\documentclass[reprint, superscriptaddress, amsmath, massymb, aps]{revtex4-2}
\usepackage[utf8]{inputenc}
\usepackage{color}
\usepackage{tocloft}
\usepackage[usenames,dvipsnames,svgnames,table]{xcolor}
\usepackage{hyperref}
\usepackage[section]{placeins} 
\usepackage{mathtools}
\usepackage{MnSymbol}
\usepackage{bm}
\usepackage{epstopdf}
\usepackage{braket}
\usepackage{wrapfig}
\usepackage{bbold}
\hypersetup{
     colorlinks   = true,
     linkcolor    = blue,
     citecolor    = blue
}
\usepackage{lipsum}
\usepackage{cancel}

\newcommand{\LG}{\mathrm{LG}}
\newcommand{\VG}{\mathrm{VG}}
\newcommand{\grad}{\nabla}

\definecolor{alexis}{rgb}{0.90, 0.1, 0.2}
\usepackage{lineno}
\begin{document}

%\title{Strong-field physics in three-dimensional topological insulators}
%\title{Theory of high nonlinear optical responses: propagation picture}
\title{Theory for all-optical responses in topological materials: the velocity gauge picture}

%\author{Denitsa Baykusheva}%
%\email{denitsab@stanford.edu}
%\affiliation{Stanford PULSE Institute, SLAC National Accelerator Laboratory, Menlo Park, California 94025, USA}
\author{Dasol Kim}
%\email{achacon@postech.ac.kr }
%\affiliation{Center for Nonlinear Studies and Theoretical Division, Los Alamos National Laboratory, Los Alamos, New Mexico 87545, USA}
\affiliation{Department of Physics and Center for Attosecond Science and Technology, POSTECH, 7 Pohang 37673, South Korea}
\affiliation{Max Planck POSTECH/KOREA Research Initiative, Pohang 37673, South Korea}

\author{Dongbin Shin}
\affiliation{Max Planck Institute for the Structure and Dynamics of Matter and Center for Free Electron Laser Science, 22761 Hamburg, Germany}

\author{Alexandra S. Landsman}
%\email{achacon@postech.ac.kr }
%\affiliation{Center for Nonlinear Studies and Theoretical Division, Los Alamos National Laboratory, Los Alamos, New Mexico 87545, USA}
\affiliation{Max Planck POSTECH/KOREA Research Initiative, Pohang 37673, South Korea}
\affiliation{Department of Physics, Ohio State University, 191 West Woodruff Ave, Columbus, OH 43210, USA}

%\author{Angel Rubio}
%\affiliation{Max Planck Institute for the Structure and Dynamics of Matter and Center for Free Electron Laser Science, 22761 Hamburg, Germany}
%\affiliation{Nano-Bio Spectroscopy Group,  Departamento de F\'isica de Materiales, Universidad del Pa\'is Vasco UPV/EHU- 20018 San Sebastián, Spain}
%\affiliation{Center for Computational Quantum Physics (CCQ), The Flatiron Institute, 162 Fifth avenue, New York NY 10010.}

\author{Dong Eon Kim}
\email{kimd@postech.ac.kr }
%\affiliation{Center for Nonlinear Studies and Theoretical Division, Los Alamos National Laboratory, Los Alamos, New Mexico 87545, USA}
\affiliation{Department of Physics and Center for Attosecond Science and Technology, POSTECH, 7 Pohang 37673, South Korea}
\affiliation{Max Planck POSTECH/KOREA Research Initiative, Pohang 37673, South Korea}

\author{Alexis Chac\'{o}n}
\email{achacon@postech.ac.kr}
%\affiliation{Center for Nonlinear Studies and Theoretical Division, Los Alamos National Laboratory, Los Alamos, New Mexico 87545, USA}
\affiliation{Department of Physics and Center for Attosecond Science and Technology, POSTECH, 7 Pohang 37673, South Korea}
\affiliation{Max Planck POSTECH/KOREA Research Initiative, Pohang 37673, South Korea}

%\author{Dasol Kim}
%\affiliation{Max Planck POSTECH/KOREA Research Initiative, Pohang 37673, South Korea}

%\author{Dong Eon Kim}
%\affiliation{Max Planck POSTECH/KOREA Research Initiative, Pohang 37673, South Korea}

%\author{Tony F. Heinz}
%\affiliation{Stanford PULSE Institute, SLAC National Accelerator Laboratory, Menlo Park, California 94025, USA}

%\author{David A. Reis}
%\affiliation{Stanford PULSE Institute, SLAC National Accelerator Laboratory, Menlo Park, California 94025, USA}

%\author{Shambhu Ghimire}
%\email{shambhu@stanford.edu}
%\affiliation{Stanford PULSE Institute, SLAC National Accelerator Laboratory, Menlo Park, California 94025, USA}

\date{\today}%
%\tableofcontents
\begin{abstract}
%Exp
%Theo 
%Ultrafast sub-cycle
%{\bf
%\noindent -- Small Intro to topology and motivtion \\ 
%-- what is the problem \\
%-- how to solve it \\
%-- results \\
%-- outlines \\
%-- main message: how to integrate the natural singularity in topological materials by means of velocity gauge?}

%We propose the laser-electromagnetic velocity gauge as a theoretical framework to integrate the Semiconductor Bloch Equations (SBEs) and calculate the High Harmonic Generation (HHG) signals in Topological Materials (TMs).~In 2D, these materials show conducting topological edge states (surface in 3D) and insulating topological bulk states, protected by symmetries and topological invariants.~

High Harmonic Generation (HHG), which has been widely used in atomic gas, has recently expanded to solids as a means to study highly nonlinear electronic response in condensed matter and produce coherent high frequency radiation with new properties.  Most recently, attention has turned to Topological Materials (TMs) and the use of HHG to characterize topological bands and invariants.  Theoretical interpretation of nonlinear electronic response in TMs, however, presents many challenges. ~In particular, the Bloch wavefunction phase of TMs has undefined points in the Brillouin Zone.~This leads to singularities in calculating the inter-band and intra-band transition dipole matrix elements of Semiconductor Bloch Equations (SBEs).~Here, we use the laser-electromagnetic velocity gauge ${\bm p}\cdot {\bf A}(t)$ to numerically integrate the SBEs and treat the singularity in the production of the electrical currents and HHG spectra.~We use a prototype of Chern Insulators (CIs), the Haldane model, to demonstrate our approach. We find good qualitative agreement of the velocity gauge compared to the length gauge and the Time-Dependent Density Functional theory in the case of topologically trivial materials such as MoS$_2$.~For velocity gauge and length gauge, our two-band Haldane model reproduces key HHG spectra features: ({\it i})~The selection rules for linear and circular light drivers, ({\it ii})~The linear cut-off law scaling and ({\it iii})~The anomalous circular dichroism.~We conclude that the velocity-gauge approach captures experimental observations and provides theoretical tools to investigate topological materials.

\end{abstract} 

\maketitle

\section{Introduction}\label{sec:intro}

%\begin{itemize}
%%\item Semi-Historical and conceptual introduction with QAHE, QSHI, QFHEs, etc. Issues in Topological materials
%%\item Ultrafast review 
%\item Issues in topological materials, still not clear
%\item Misha approach
%\item Chacon approach 
%\item advantage and disadvantage 
%\item hypothesis of this paper or proposal
%\end{itemize}

%{\color{red} \textbf{List of figures modification}
%\begin{itemize}
%    \item Noisy figure from LG-Hamiltonian
%    \item vector field with Bconnection
    %\item change font sizes for figures, and add (a) or (b)
    %\item Add lattice structure + laser %polarization subset
    %\item Try to improve GK comparison figure - %more matching
    %\item add line in cutoff scan
    %\item conduction population timescan figure
    %\item sum-rule visualization for each H %components
    %\item using LG dephasing in VG and see the %results
    %\item population and other observable also %calculate once more with LG dephasing
%    \item change T2 order in T2scan graph
%\end{itemize}
%}

\begin{figure}[htbp]
\begin{center}
\includegraphics[width=0.42\textwidth]{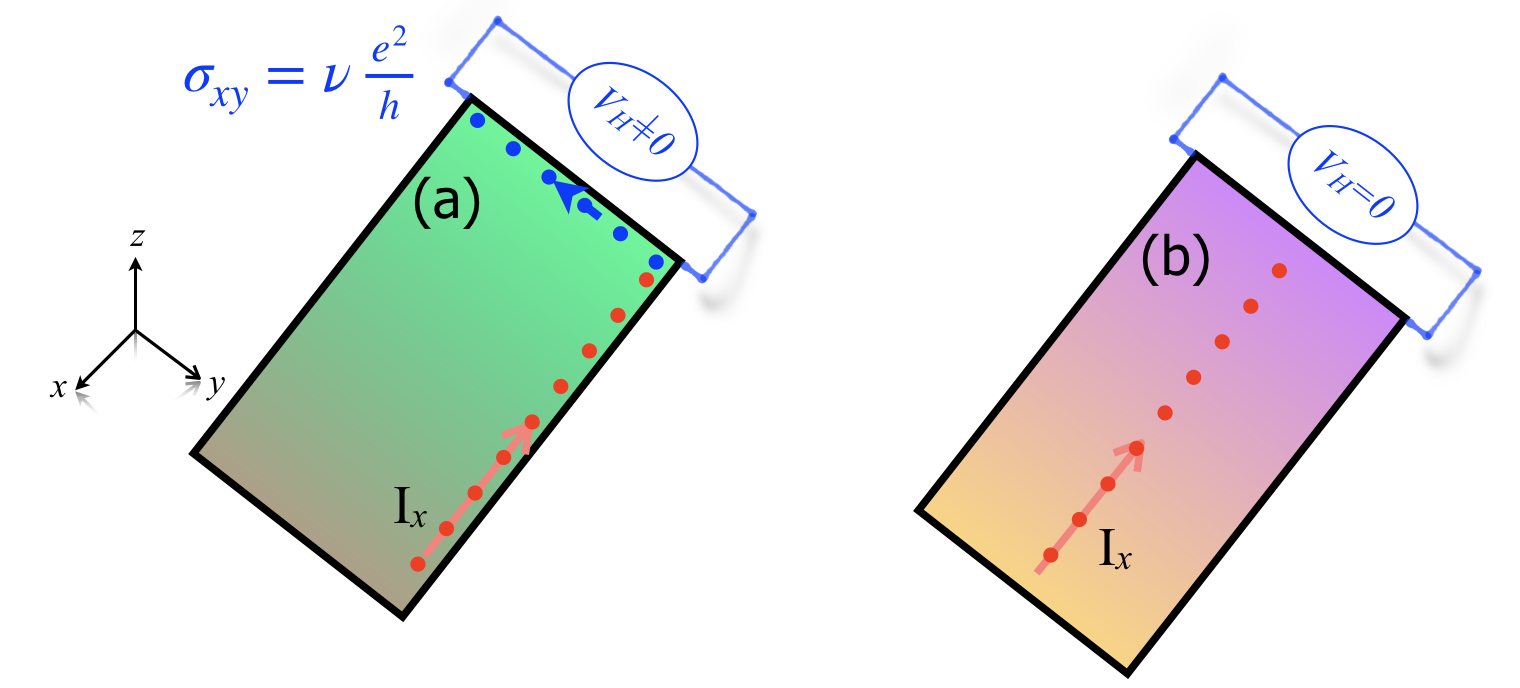}
\caption{Quantum Anomalous Hall Effect. Panel (a) depicts a cartoon of the longitudinal electrical currents $I_x$ (red dots) and the {\it quantized transversal conductivity} $\sigma_{xy}=\nu\frac{e^2}{h}$ (Hall voltage $V_{H}\neq0$ or charge current in blue dots) in an ideal Chern Insulator~(CI) proposed by Haldane~\cite{Haldane1988} (the Chern Number or topological invariant can be $\nu=\pm1$), i.e.~the Quantum Hall Effect without any applied external magnetic field.~Panel (b), shows an example of traditional conductors which do not exhibit any quantized transversal Hall voltage ($V_{H}=0$ or conductivity), $\nu=0$.}
\label{fig:Figure0}
\end{center}
\end{figure}

The measurement of the Quantum Spin Hall Effects (QSHEs)~\cite{konig2007,bernevig2006,Zhang2009,Chen2009,OliaeiMotlagh2018} in HgTe (2007) ushered a new era of condensed matter physics, paving the way for unexpected technological advances~\cite{Hasan2010,Zhang2009,ShouCheng2011}.~The HgTe material consists of quantum wells that exhibit transversal spin currents at the edge, but insulating features in the bulk under a static longitudinal voltage~\cite{Xiao2010,Hasan2010}~(see Fig.~\ref{fig:Figure0}).~These edge currents and insulating bulk suggest unique applications for TIs, including in metrology ~\cite{Haddad2016} and the control of quantum logic operations~\cite{Kung4006}.
%~Mathematically, these TMs are defined by the~Bloch wavefunction~\cite{Hasan2010,Xiao2010}, $| \Psi^s_{m,{\bm k}} \rangle = \frac{1}{\rm \sqrt{N}} \exp{(i{{\bf r}\cdot {\bm k}})}|u^s_{m,{\bm k}}\rangle$, (${\it i}$) here $|u^s_{m,{\bm k}}\rangle$ is the periodic part of the Bloch states for the m$^{th}$-band of energy $\varepsilon_m({\bm k})$ in the Brillouin Zone (BZ)~\cite{konig2007,bernevig2006}, $s$ represents the spin-index~\cite{Kane2005,Hasan2010}, ($\it ii$)~the symmetries of the Hamiltonian $\hat{H}_0({\bm k})$,  and $(\it iii)$~{\it the topological invariant.}~

The quantum wells of HgTe, as well as other materials such as CdTe~\cite{bernevig2006}, have topological invariants belonging to the class ${\mathbb Z}_2$ of TIs. This is defined in terms of the wavefunction parities or Berry phase~\cite{bernevig2006,Hasan2010}.
This 2D TIs is a unique phase of matter in the sense that the edge (surface) current is protected by the {\it time-reversal symmetry} of the Hamiltonian and its {\it topological invariant}~\cite{Kane2005}. This symmetry protects against dissipation and provides robustness against perturbations of the topological materials~\cite{KosterlitzNobel2016}.

Despite widespread interest in nonlinear interaction of topological materials with ultrafast laser pulses, there is little research on the topic due to the difficulties in theoretical modeling and interpretation of resulting higher harmonic emission. Addressing these challenges is instrumental to guiding future experimental observations in TMs. In this paper, we expand the study of non-linear optical emission to TMs by solving the SBEs in the laser-electromagnetic velocity gauge (VG)~${\bm p}\cdot {\bf A}(t)$ in the mid-infrared (MIR) laser regime.
In particular, we introduce a new approach to address the integration of singularity in the SBEs for the highly non-linear optical response in the Topological Materials (TMs).

There is a wide variety of TMs depending on the {\it topological invariant} and the Hamiltonian symmetries of these materials.~These classifications are organized in the periodic table of TMs (Insulators), shown in Ref.~\cite{Hasan2010}.~Depending on the dimensionality of the samples, symmetries, and topological invariant, this table shows TMs with charge currents, locked-spin up and down currents~\cite{Shin2019,Haldane1988}, and Weyl fermions, among others. For instance,~ the QSHE leads to~QSH insulators HgTe (2D TI) or Bi$_2$Se$_3$ (3D TI).

%

%Nevertheless, these are not the only type of materials belonging to the periodic table of TMs~\cite{Hasan2010} what exhibit Quantum Anomalous Hall Effects (QAHEs). 
Haldane in~1988 introduced the first paradigmatic class of TMs that shows Quantum Anomalous Hall Effects (QAHEs) (see Appendix~\ref{sec:Haldane}).~The Haldane model exhibits {\it quantized conductivities}, $\sigma_{xy}= \nu\, \frac{e^2}{h}$ at the edge, where $e$ is the electron charge, $h$, the Plank constant and $\nu$, the topological invariant (or the Chern Number). This invariant is a quantized integer number that characterizes topological CIs (see Fig.~\ref{fig:Figure0}).~The topological states have {\it singularities} in the BZ, which can lead to numerical problems (see Fig.~2~of Ref.~\cite{Chacon2020}).~Here, we treat these {\it singularities} by using the laser-electromagnetic velocity gauge (VG). As proof of concept, we use the Haldane model to test and validate this approach.

\begin{figure}[htbp]
\begin{center}
\hspace{-0.1cm}\includegraphics[width=0.35\textwidth]{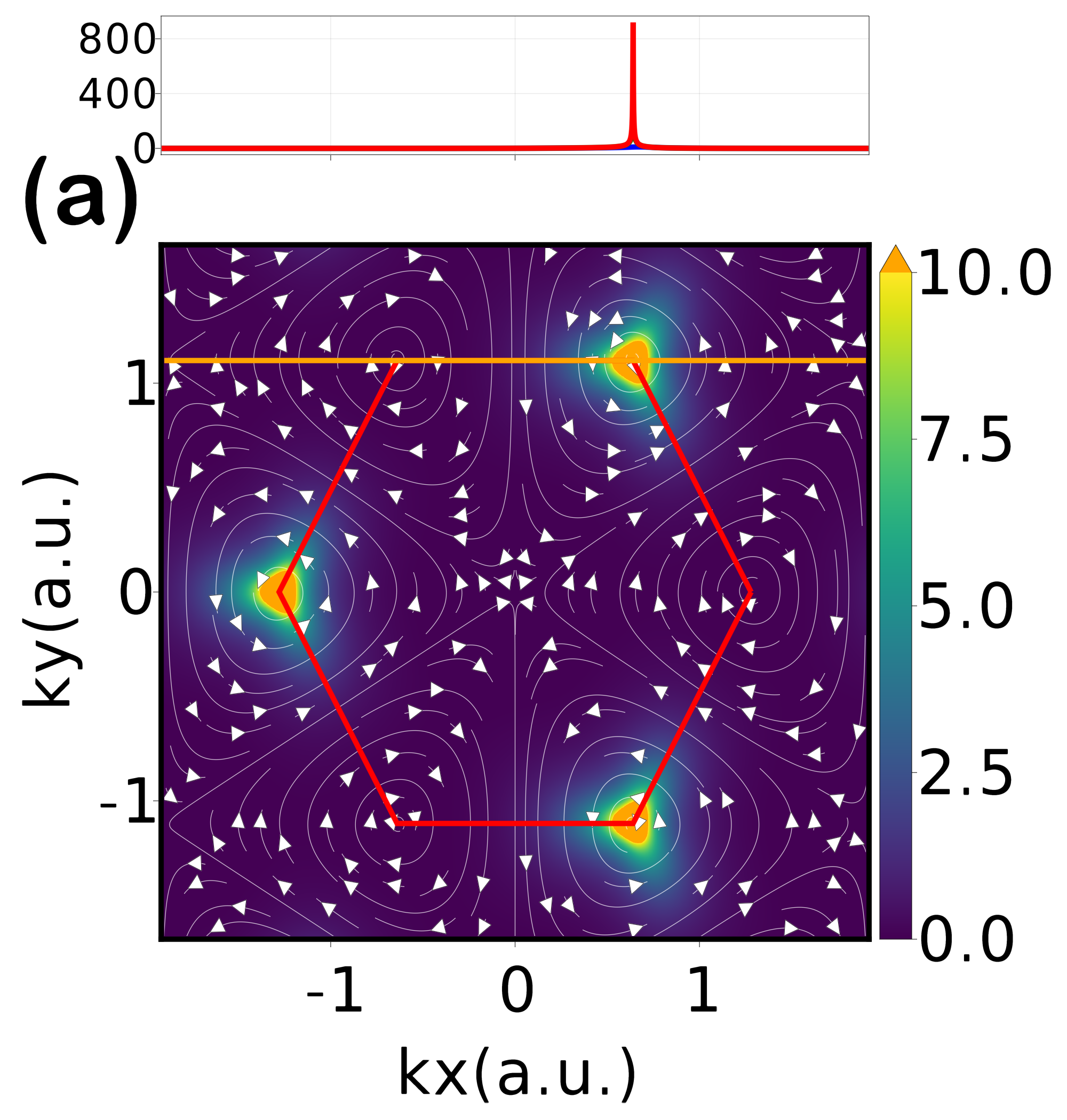}\\ 
\includegraphics[width=0.35\textwidth]{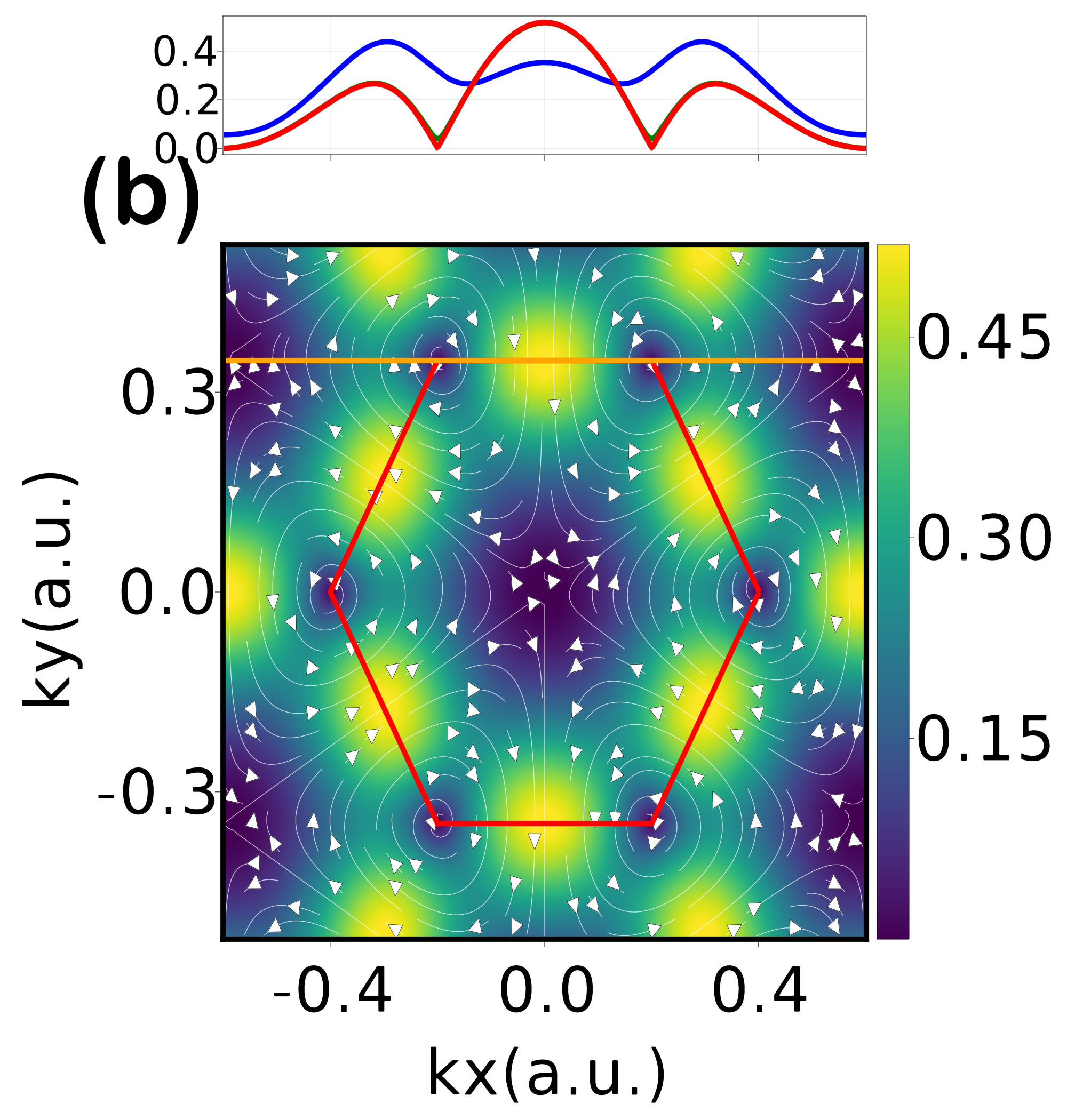}
\caption{Singularity in the Berry connection of topological material. (a) The absolute value of the Berry Connection for Chern insulators defined in the topological Haldane model and (b) same as in (a) but for a topologically trivial material, $\mathrm{MoS}_{2}$. The vectorial field indicates how the Berry connection can accumulate a phase in the Haldane model. In other words, by the Stokes' Theorem, the topological Chern Number is $\nu_n=\frac{1}{2\pi}\oint_{C}{\bm\xi}_n({\bm k})\cdot d{\bm k}$, where $C$ denotes a closed line-path integral.~Upper panels show a 1D cross-section along the orange line for trivial and topological phase, i.e., the absolute value of Berry connection for small $k_{y}$-offsets, respectively.}
\label{fig:Figure1}
\end{center}
\end{figure}

Figure~\ref{fig:Figure0}(a) depicts the Quantum Hall Effect (QHE) without Landau levels~\cite{Haldane1988,Hasan2010,KosterlitzNobel2016}.~This shows that $\nu$ is essential for topological materials.~Moreover, this critical aspect of TMs is contained in the undefined phase of the topological states. The singularity itself is independent of the wavefunction gauge, $|{\tilde u}_m^s\rangle=\exp{(i\phi_m)}|{\tilde u}_m^s\rangle$: the $\bm k$-position of the singular point can be manipulated via the gauge transformations, but no eliminated in TMs.~The latter leads to an interconnection of the singularity in the wavefunction phase with $\nu$~\cite{Kohmoto1985}. Kohmoto showed that {\it without this singularity, no {\rm QHEs} is observed}~\cite{Kohmoto1985} and the material behaves as an ordinary semiconductor or conductor (see Fig.~\ref{fig:Figure0}(b)). 
Unfortunately, this singularity of the wavefunction affects the calculation of the Chern number which is defined by~\cite{Haldane1988,Hasan2010,KosterlitzNobel2016,Shin2019}:
\begin{eqnarray}
\nu_m = \frac{1}{2\pi}\int_{\rm BZ} d^2{\bm k}\cdot {\bm \Omega}_m({\bm k}),
\end{eqnarray}
where ${\bm \Omega}_m({\bm k})=\langle \partial_{\bm k}u_{m,{\bm k}}| \times |  \partial_{\bm k}u_{m,{\bm k}}\rangle=\nabla_{\bm k}\times{\bm\xi}_m({\bm k})$ is the Berry curvature, the Berry connection, ${\bm\xi}_m({\bm k})=i\langle u_{m,{\bm k}} |\nabla_{\bm k} u_{m,{\bm k}}\rangle $ and the transition dipole matrix elements, ${\bm d}_{mn}({\bm k})=i\langle u_{m,{\bm k}} |\nabla_{\bm k} u_{n,{\bm k}}\rangle $. Hence, this singularity extremely complicates the calculations of the dipoles and Berry connections (see Ref.~\cite{Chacon2020} and Fig.~\ref{fig:Figure1}(a) in comparison to Fig.~\ref{fig:Figure1}(b)).

On the other hand, the ultrafast non-linear optical spectroscopy and High Harmonic Generation (HHG) in topological materials are attracting the attention of ultrafast physics and condensed matter communities~\cite{Bauer2018,Drueke2019}.~This non-linear optical spectroscopy explores how {\it topological invariants} are encoded in the high harmonic spectrum, and is a complementary alternative to Angle-Resolved Photoelectron measurements ~\cite{DenitsaPRA2021,Bauer2018,Cavalleri2020,Bauer2018}. However, the use of HHG to characterize TMs is very much in its infancy. ~A few recent studies of HHG in the paradigmatic Haldane model~\cite{Silva2019,Misha2018,Bauer2018} have shown the complexity of computing the non-linear currents using the SBEs. 

The evolutionary density matrix ${\hat\rho}({\bf K},t)$ or Semiconductor Bloch equations (SBEs) reads: 
\begin{eqnarray}
    \frac{\partial}{\partial t}{\hat \rho}_{mn}({\bf K}, t) &= -i\left[ \varepsilon_{mn}({\bf K} + {\bf A}(t))-\frac{i}{T_2}\right]{\hat \rho}_{mn}({\bf K}, t) \nonumber \\
    &-i{\bf E}(t)\cdot\left[\bm{D}({\bf K} + {\bf A}(t)), {\hat \rho}({\bf K}, t) \right]_{mn}. \label{eq:SBEs_lg}
\end{eqnarray}
The above equation contains the singular term of the topological wavefunction: the Berry connection and the dipole matrix element which are encoded in ${\bm D}_{mn}({\bm k})$~\cite{Chacon2020} (see below for mathematical definition).~This ${\bm D}_{mn}({\bm k})$ contains both the inter-band dipole matrix element ${\bf d}_{mn}({\bm k})$ for $m\neq n$, and the intra-band Berry connection ${\bm\xi}_m({\bm k})$ for $m=n$.~Here ${\bf A}(t) = -\partial_t {\bf E}(t)$ is the vector potential of the electric field ${\bf E}(t)$, and the energy difference between the $m^{\rm th}$ and  $n^{\rm th}$ bands,~$\varepsilon_{mn}({\bm k})=\varepsilon_{m}({\bm k})-\varepsilon_{n}({\bm k})$. We use the so called moving BZ frame~\cite{VampaPRL2014}, ${\bm k} = {\bf K}+{\bf A}(t)$, and the phenomenological dephasing time $T_2$. The singularities in ${\bm D}_{mn}({\bm k})$ in Eq.~(\ref{eq:SBEs_lg}) induces numerical errors in the calculation of high-order harmonics from TMs, more noticeable in strong field regimes. These lead to {\it wrong plateau} and {\it cut-off structures} of the HHG spectra (see Ref.~\cite{Chacon2020}), if the singular integral in Eq.~(\ref{eq:SBEs_lg}) is not handled properly.

To address this problem, we previously developed ({\it a}) the variable ``matter-gauge wavefunctions" method, which considers the pseudo-spin gauge Hamiltonian and the periodicity of the Haldane model in the BZ~\cite{Chacon2020}.~This method showed an excellent resolution of the harmonic orders (HOs) and cut-off of the HHG spectra. Additionally, theoretical efforts by Silva et al.~\cite{SilvaPRB2019} handled this singularity by using ({\it b}) the time-evolution of $\hat\rho({\bf k},t)$ in the Maximally Localized Wannier basis (MLWB).

Each method, either ({\it a}) or ({\it b}) has its advantages and disadvantages. For instance, in Ref.~\cite{Chacon2020}, method ({\it a}) only works in the case of Tight Binding Approximations (TBAs).  In Ref.~\cite{SilvaPRB2019}, method ({\it b}) has the disadvantage in the evaluation of the dephasing time at each time-step $t$. The evaluation of $T_2$ makes its numerical implementation tedious for straightforward technical development compared to the VG. The MLWB method requires evaluations of $T_2$ in the Hamiltonian-gauge instead of its original Hamiltonian-Wannier-gauge representation (increasing the number of computational operations). Note, we have verified that both methods ({\it a}) and ({\it b}) reach the same results in the Hamiltonian Bloch basis.

Theoretical study on the laser-electromagnetic gauge symmetry in trivial materials can be found in Ref.~\cite{PeresPRB2011}. This prominent study found gauge invariance of the non-linear optical responses only under a specific number of bands. The truncation of the SBEs solution as a function of number of bands breaks the gauge-symmetry~\cite{PeresPRB2011} in the calculation of the charge currents (see Appendix~\ref{sec:gauge}).
%Appendix B

%In this paper, we introduce a new approach to address the integration of this singularity in the SBEs for the high non-linear optical responses from Topological Materials (TMs).
%SASHA:  I moved this up to the first page.  I think this is the 3rd or 4th time we are saying "In this paper we did blah..." in different places in the manuscript, so I wanted to consolidate that.

We introduce the laser-electromagnetic velocity gauge to compute the SBEs and the electrical currents in trivial materials and TMs. Furthermore, we compare the HHG spectra produced by the proposed VG with the length gauge (LG).  The rest of the paper is organized as follows:
In Section~(\ref{sec:TF00}), we describe the electrical current and derive the SBEs in the laser-electromagnetic length gauge and velocity gauge. 
In section (\ref{sec:NRV00}), we use the VG to calculate high harmonic emission from trivial material such as a monolayer of MoS$_2$. 
To confirm the validity of our approach, we compare these results with the LG and the Time-Dependent Density Functional Theory (TDDFT) for MoS$_2$. 
We then extend the application of the proposed VG theoretical framework to topological Chern Insulators~(CIs) for both linearly and circularly polarized MIR or THz light sources. 
Our approach is further validated by computing the cut-off law~\cite{GhimireNatPhy2011} of the HHG spectra and the Circular Dichroism in topological materials. We also compare these outcomes with the length gauge results. 
In Section~(\ref{sec:D00}), we discuss the advantages and disadvantages of LG and VG pictures in computing the HHG spectra and conclude that VG is a suitable and straight-forward method to calculate the HHG spectra from topological materials.

\section{Theoretical Framework}\label{sec:TF00}
\noindent In the dipolar approximation, the velocity gauge, and length gauge are theoretically used to describe the non-linear optical responses~\cite{virk2007,kruchinin2013,yakovlev2017} from solids subjected to ultrashort lasers.~The total Hamiltonian of the laser-lattice system is ${\hat H}(t)={\hat H}_0 + {\hat V}_{\rm int}(t)$, the interaction term~${\hat V}_{\rm int}(t)$ in the VG or LG theoretical framework reads: 
 
\begin{eqnarray}
{\hat V}_{\rm int}^{\rm (VG)} &=& {\hat{\bm p}}\cdot {\bf A}(t) + {\bf A}^2(t)/2,\,\,{\rm and}\,\label{eq:VGeq}\\
{\hat V}_{\rm int}^{\rm (LG)} &=& {\hat{\bm x}}\cdot {\bf E}(t).\,\,\label{eq:LGeq}
\end{eqnarray}
Although these laser-electromagnetic gauges should provide the same results for a physical observable, for instance, the HHG spectra, previous studies have found that, unfortunately, it is not the case in several systems~\cite{yakovlev2017,Granados2012,JAHAndPlaja2009}.~This breaking of the gauge-symmetry occurs when an approximation is carried out to solve the Time-Dependent Schr\"odinger Equation (TDSE). Note, however, that the full numerical integration of the TDSE for the HHG spectra under the LG and VG is covariant~\cite{ChaconPRA2015,Cormier1996}.  This supports the observation that any approximation of the TDSE can break the laser-electromagnetic gauge-symmetry. For instance, in the Strong Field Approximation (SFA) applied to a gas, this laser-electromagnetic gauge-symmetry is broken~\cite{MaciekROOP2019}; producing different HHG spectra, particularly for the emitted intensity yield~\cite{Granados2012,JAHAndPlaja2009}. Nevertheless, the main qualitative features of the high harmonics are reproduced by the LG and VG in the SFA formalism. ~In the case of TMs, we expect a similar trend.

%%%%%%Apr 28, 2021%%%%%%%%%%
%Description of the problem LG 

In the length gauge, the numerical integration of the SBEs is an extremely problematic task: the ${\bm k}$-space position operator depends on the crystalline momentum derivatives $i\partial_{\bm k}$ in the BZ (see Eq.~(\ref{eq:x_Bloch})). Numerically, this finite ${\bm k}_j$-neighbor will couple non-define SBEs, including the electronic density and electron coherence of the ${\hat \rho}({\bm k},t)$ operator, near the singular point of the Berry connection and dipoles. This problem has been found not only in topological materials but also in trivial~Ref.~\cite{YueGaarde2020}. 

~On the contrary, the velocity gauge offers a way to avoid the singularity issue by the de-coupling of the neighboring ${\bm k}_j$.~Hence, notwithstanding some disadvantages of the VG, related to the Bloch acceleration theorems~\cite{virk2007,kruchinin2013,yakovlev2017} and crystal kinetic momenta, it offers an attractive alternative to the typically used length gauge.  
%. However, the cost of modeling an approximated HHG spectrum in TM is mainly in favor of the VG compared to the acceleration theorem. In this paper, 
We therefore propose the velocity gauge as an alternative to the LG to study the non-linear optical responses from topological materials.

\subsection{Velocity gauge picture}\label{sec:VGP00}

\noindent Commonly, in a periodic crystalline structure subjected to an external laser-field, the charge current is calculated by integrating the ${\bm k}$-elementary-microscopic currents in the BZ: 
\begin{eqnarray}
    \bm{J}(t) = \int_{\text{BZ}} \frac{d^{3}k}{(2\pi)^{3}} \bm{j}(\bm{k}, t). \label{eq:current0}
\end{eqnarray}
Here we define the elementary-microscopic current ${\bm j}({\bm k},t)$ as
\begin{eqnarray}
    \bm{j}(\bm{k}, t) &=& -\text{Tr}\left({\hat\rho} \hat{\bm{v}}\right) \nonumber\\
    &=& \text{Tr}\left({\hat\rho}(\bm{k}, t) {\hat{\bm p}}\right) -N_{\text{VB}}\bm{A}(t) \nonumber\\
    &=& -\sum_{m,n}\rho_{mn}(\bm{k}, t)\bm{P}_{nm}(\bm{k}) -N_{\text{VB}}\bm{A}(t). \label{eq:current1}
\end{eqnarray}
This corresponds to the expectation value of the {\it velocity operator} ${\hat{\bm v}}=-i\left[{\hat H}(t),\hat{\bm x} \right]$.~The current $ \bm{j}(\bm{k}, t)$ is defined in terms of the density matrix $\hat{\rho}=\hat{\rho}(\bm{k}, t)$, the momentum matrix element $\bm{P}$, and the number of valence band $N_{\text{VB}}$~\cite{YueGaarde2020,kruchinin2013,yakovlev2017}. 

The time-propagation of the density matrix ${\hat \rho}({\bm k},t)$ is given by Liouville-von Neumann equation,
\begin{eqnarray}
    i\frac{\partial {\hat \rho(\bm{k}, t)}}{\partial t} = \left[ \hat{H}(t), {\hat \rho(\bm{k}, t)} \right], \label{eq:Liouville}
\end{eqnarray}
where $\hat{\rho}$ will be evaluated in the VG via ${\hat H}(t)$. 

\subsubsection{Hamiltonian representation in the VG}
\noindent Usually, the Hamiltonian representation is defined by the Bloch states for the laser-free Hamiltonian, ${\hat H}_0$. In the VG, the Hamiltonian describing the laser-periodic crystalline interaction reads:
\begin{eqnarray}
    \hat{H}(t) = \hat{H}_{0} + \hat{\bm{p}}\cdot{\bf A}(t). \label{eq:Hamiltonian_vg}
\end{eqnarray}
We neglect the term proportional to $A^{2}(t)$ of the interacting ${\hat V}^{\rm  (VG)}_{\rm int}({\bm k},t)$ of Eq.~(\ref{eq:VGeq}) (for details, see Ref.~\cite{kruchinin2013}). Thus, the time-momentum evolution of the density matrix elements in the VG reads, 
\begin{eqnarray}
\dot{{\rho}}_{mn}({\bm k}, t) = -i\left[\varepsilon_{mn}(\bm{k}) - i \frac{1}{T_{2}}\right]\rho_{mn}(\bm{k}, t)  \nonumber \\
    \hspace{-.1cm}- i{\bf A}(t)\cdot \sum_{l} \left[\bm{P}_{ml}({\bm k})\rho_{ln}({\bm k}, t) - \bm{P}_{ln}({\bm k})\rho_{ml}({\bm k}, t) \right]. \label{eq:SBEs_vg}
\end{eqnarray}
Here $\varepsilon_{mn}({\bm k})$ is the energy difference between the band $m$ and $n$, with the phenomenological dephasing time given by $T_{2}$.

The advantage of the VG is that every ${\bm k}$-crystal momentum channel is de-coupled~\cite{kruchinin2013,yakovlev2017}. Hence one can choose a discretized ${\bm k}$-grid that avoids the singularity.~Additionally, we can quickly parallelize the implementation of the code for Eq.~(\ref{eq:SBEs_vg}). For instance, we use Message Passing Interface (MPI) in ${\rm C++}$ and numerically solve Eq.~(\ref{eq:SBEs_vg}) using Runge-Kutta $5^{\text{th}}$ order method.

\subsection{Length gauge pictures}\label{sec:LGP00}
\noindent The evolution of the electronic density operator~${\dot{\hat \rho}}({\bm k},t)$~in length gauge and the ``Hamiltonian matter-gauge'' can be acquired in a similar procedure as described in Refs.~\cite{kruchinin2013,yakovlev2017,Chacon2020}. From Liouville-von Neumann equation given by Eq.~(\ref{eq:Liouville}), considering  the interacting potential of Eq.~(\ref{eq:LGeq}) and the position operator in the Bloch basis $\hat{\bm{x}}$~\cite{Blount1962}:
\begin{eqnarray}
    \hat{\bm{x}}_{mn} = \left(-i{\bm\nabla}_{\bm k} + {\bm \xi}_{m}\right)\delta_{mn} + \bm{d}_{mn}, \label{eq:x_Bloch}
\end{eqnarray}
${\dot{\hat \rho}}({\bm k},t)$ can be expressed as Eq.~(\ref{eq:SBEs_lg}).~This representation is sensitive to the singularity of the topological states in TMs, as already discussed above. The position operator indeed contains intra-band momentum terms, which are defined in ${\bm \nabla}_{\bm k}$. In a finite and a discretized ${\bm k}$-space grid, this means that the ${\hat \rho}({\bm k},t)$ depends on its ${\bm k}$-space ``numerical neighbour cell" and the electric field strength.~This is problematic in LG and in its Hamiltonian representation. Even in the case that one can express the time-evolution of ${{\hat \rho}}({\bm K},t)$ in terms of the moving frame ${\bm k}= {\bm K} + {\bf A}(t)$~\cite{VampaPRL2014,VampaJPB2017,Chacon2020}, the vector potential ${\bf A}(t)$ will force the ${\dot{\hat \rho}}({\bm K},t)$ to travel throughout the singularity described in Fig.~\ref{fig:Figure1}(a) for TMs.

\subsubsection{Wannier representation for the LG}
\noindent The numerical solution to Eq.~(\ref{eq:SBEs_lg}) requires continuous quantities such as transition dipole matrix elements and Berry connections. Unfortunately, this is not possible for topological materials~\cite{Kohmoto1985,Chacon2020}.~The Wannier representation~\cite{silva2019a} promises to address this problem. From the application of Eq.~(\ref{eq:Liouville}), considering TBA as a basis and Eq.~(\ref{eq:SBEs_lg}), the density equation of motion in the Wannier basis yields~\cite{silva2019a}:
\begin{eqnarray}
    i\frac{\partial }{\partial t}{\hat\rho}^{\rm (W)}(\bm{K}, t) = \left[ {\hat H}_{0}^{\rm{(W)}}(\bm{K}+\bm{A}(t)), {\hat\rho}^{\rm (W)}(\bm{K}, t) \right] \nonumber \\
    + {\bf E}(t)\cdot\left[\bm{D}^{\rm (W)}(\bm{K}+{\bm A}(t)), {\hat\rho}^{\rm (W)}(\bm{K},t) \right]. \label{eq:SBEs_Wannier}
\end{eqnarray}
Here, ${\hat H}^{\rm (W)}_{0}({\bm k})$ is expressed in the TBA Hamiltonian.~$\bm{D}^{\rm (W)}(\bm{k})$ and ${\hat\rho}^{\rm (W)}(\bm{k}, t)$ are dipole matrix and density matrix in Wannier basis. $\bm{D}^{\rm (W)}(\bm{k})$ is calculated by 
\begin{eqnarray}
    \bm{D}_{nm}^{\rm (W)}(\bm{R}) = \sum_{\bm{R}} e^{i\bm{k}\cdot\bm{R}} \braket{\bm{0}n|\hat{\bm{r}}|\bm{R}m}. \label{eq:dipole_Wannier}
\end{eqnarray}
This exhibits continuous dipoles even in case of topological materials. Furthermore, if we assume that $\bm{D}^{\rm (W)}({\bm k})$ is diagonal, for instance:
\begin{eqnarray}
    \braket{\bm{0}n|\hat{\bm{r}}|\bm{R}m} = \delta_{\bm{0R}}\delta_{nm}{\bm \Delta}_{n},
\end{eqnarray}
one can treat $\bm{D}^{\rm (W)}(\bm{k})$ as a ${\bm k}$-independent term~\cite{VanderbiltPRB2006}. Here ${\bm \Delta}_{n} = \braket{\bm{0}n|\hat{\bm{r}}|\bm{0}n}$ is the center of $n^{\mathrm{th}}$ Wannier function or can be understood as a position of the corresponding atomic orbital. 

\section{Numerical results and validations}\label{sec:NRV00}
\noindent Our velocity gauge approach is first validated in a topologically trivial material. Subsequently, we extend the VG approach to the paradigmatic Haldane model.~In the case of trivial materials, we also calculate the HHG spectra in the length gauge and the TDDFT ~\cite{AngelOctopus2020}, and then compare it to the VG results. 
%Thus we compare our HHG outcomes from VG to LG and TDDFT in a monolayer of MoS$_2$. 
For the trivial material, we use a simplified two-bands TBA of monolayer MoS$_2$. In other words, we adjust the TBA parameters to reproduce the minimum energy gap and maximum energy gap of~MoS$_2$ (for details, see Appendix~\ref{sec:Haldane}).  We simulate the HHG spectrum using VG and LG via Eqs.~(\ref{eq:SBEs_vg}) and~(\ref{eq:SBEs_Wannier}), respectively.

\begin{figure}
    \centering
    \includegraphics[width=0.48\textwidth]{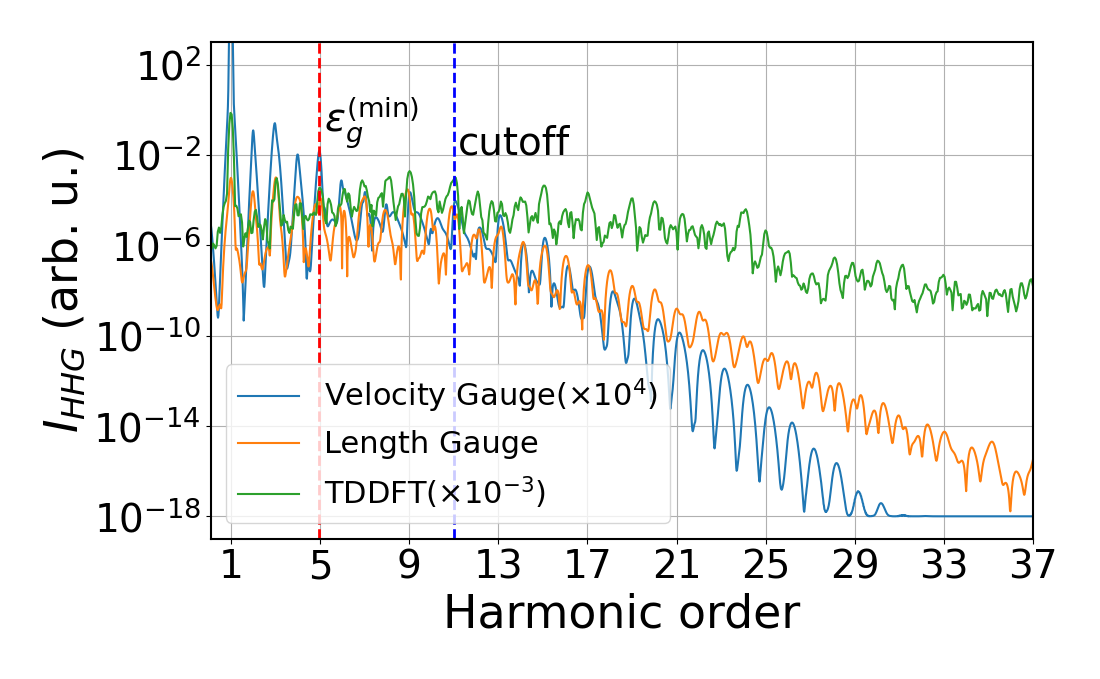}
    \caption{Calculated HHG spectra from $\mathrm{MoS}_{2}$ by the SBEs in the length gauge (orange line), velocity gauge (blue line) and TDDFT (green line). Our laser parameters are  $\hbar\omega_{0}$ = 0.3626 eV, peak electric field $E_{0}$ = 0.01265~a.u. and time pulse duration at Full-Width at Half Maximum (FWHM) of 14~opt.~cycles under a gaussian envelope, with dephasing time $T_{2}=2.7$ fs. To mimic the band structure of $\mathrm{MoS}_{2}$, we used $M_{0}$ = 0.9 eV = 0.0331 a.u., $t_{1}$ = 0.4 eV = 0.0147 a.u., $t_{2}$ = 0.667 eV = 0.0245 a.u., and $\phi_{0}$ = 0 rad for Haldane model parameters. Red dotted line indicates band gap of the $\mathrm{MoS}_{2}$ which is 1.8~eV. }
    \label{fig:TrivialGK}
\end{figure}
%

%We simulate the HHG spectrum using VG and LG via Eqs.~(\ref{eq:SBEs_vg}) and~(\ref{eq:SBEs_Wannier}), respectively. 
%~We refer LG in the Wannier representation as LG unless otherwise mentioned. 

\noindent Figure~\ref{fig:TrivialGK} shows HHG signals for both gauges in the trivial phase of $\mathrm{MoS}_{2}$.~The spectrum produced by the VG is qualitatively similar in essential features to the LG and the TDDFT calculations. For example, the VG plateau with even and odd HO structures and cut-off have a good qualitative agreement with the other two methods. For better visualization, all three calculations are normalized to have similar low-order harmonics yields.

%VG results are multiplied by a factor of four orders of magnitude and the~TDDFT results are multiplied by a factor of three orders of magnitude.

The LG and VG will yield identical results ~\cite{han2010}, if and only if the full eigenstates and eigen energies of the Hamiltonian $H^{\rm (W)}_0$ are considered in the simulation and the sum rule in Appendix~\ref{sec:gauge} is satisfied. Moreover, since we use TBA up to two states and the second nearest neighbor hopping for the Hamiltonian $H^{\rm (W)}_0$, the HHG calculations can break the laser-electromagnetic gauge-symmetry. This effect is similar to the HHG in gases.

The difference between both gauges in comparison to the TDDFT can be explained by several factors, such as the incomplete sum rule between position and momentum operator in Eq.~(\ref{eq:commutator_rp}) of Appendix~(\ref{sec:gauge}).~The incomplete basis set of TBA breaks this commutation relation, and of course the gauge-symmetry too.
For solids described in the plane-wave basis, it has been proven that the VG requires up to the 30$^{\rm th}$ band to obtain convergence~\cite{yue2020}, compared to two-bands in LG. ~Another origin of the difference between the VG and the LG is the action of the dephasing time $T_{2}$ in ${\hat\rho}({\bm k},t)$.~This phenomenological variable plays a different role in the two gauges (for details, see Appendix~\ref{sec:dephasingSec}). Note, however, that our HHG spectra show a similar tendency in both gauges; for example, the plateau structure and cut-off are similar in the VG and LG (See Fig.~\ref{fig:TrivialGK}). 

\begin{figure}
	\includegraphics[width=0.48\textwidth]{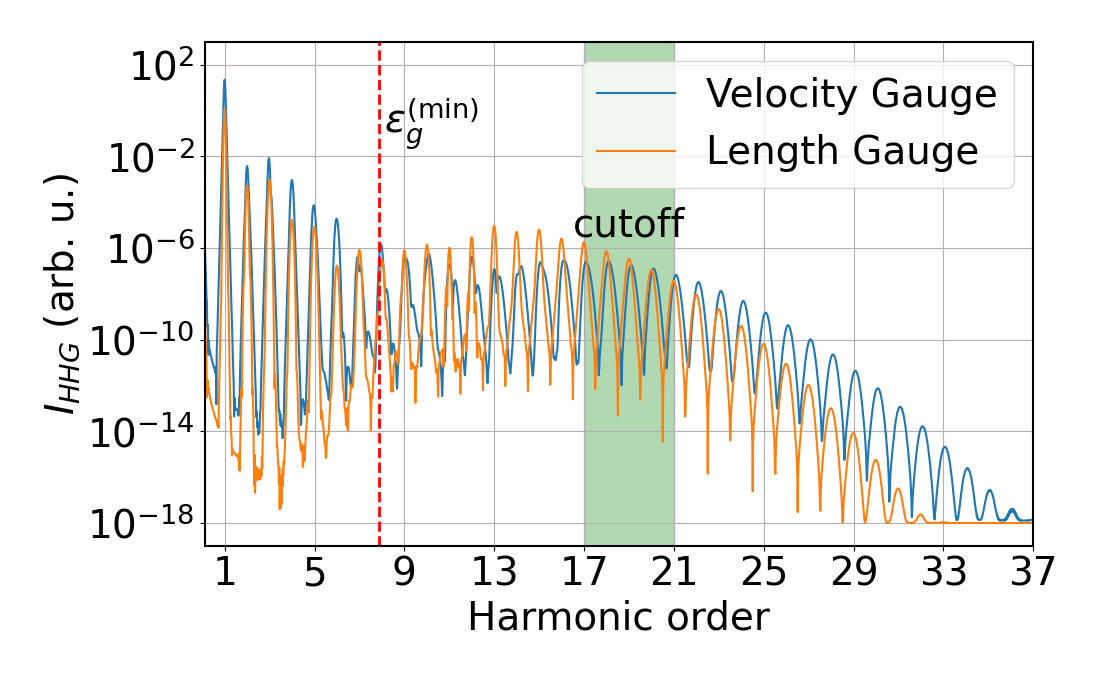}
	\caption{Calculated HHG spectra from Chern insulator in the length and velocity gauges. Laser central frequency $\omega_{0} = 0.38$~eV, peak electric field $E_{0} = 0.0045$~a.u. and FWHM duration of 14 cycles under a gaussian envelope are used. For dephasing time $T_{2}=5.3$~fs is used. The Chern insulator has $M_{0} = 0.0635$~a.u., $t_{1} = 0.075$~a.u., $t_{2} = 0.025$~a.u., and $\phi_{0} = 1.16$~rad for Haldane model parameters.~Red dotted line indicates band gap of material which is $3.0$~eV.}
	\label{fig:TopoGK}
\end{figure}

\subsection{Topological nonlinear optical response: the velocity gauge}\label{sec:TopoResults}

We now extend the VG model to topological materials.~The Haldane model (HM) belongs to the first class of topological Chern Insulators (CIs). We use the topological HM to study non-linear optical emissions and charge currents induced by the laser-CI interactions. This prototype of CI will test our velocity gauge approximation in TMs by comparing our HHG simulations in the VG to the LG. 

%%Apr 29, 19:28h

\begin{figure*}
    %\centering
    \begin{tabular}{c c}
        \includegraphics[width=0.4\textwidth]{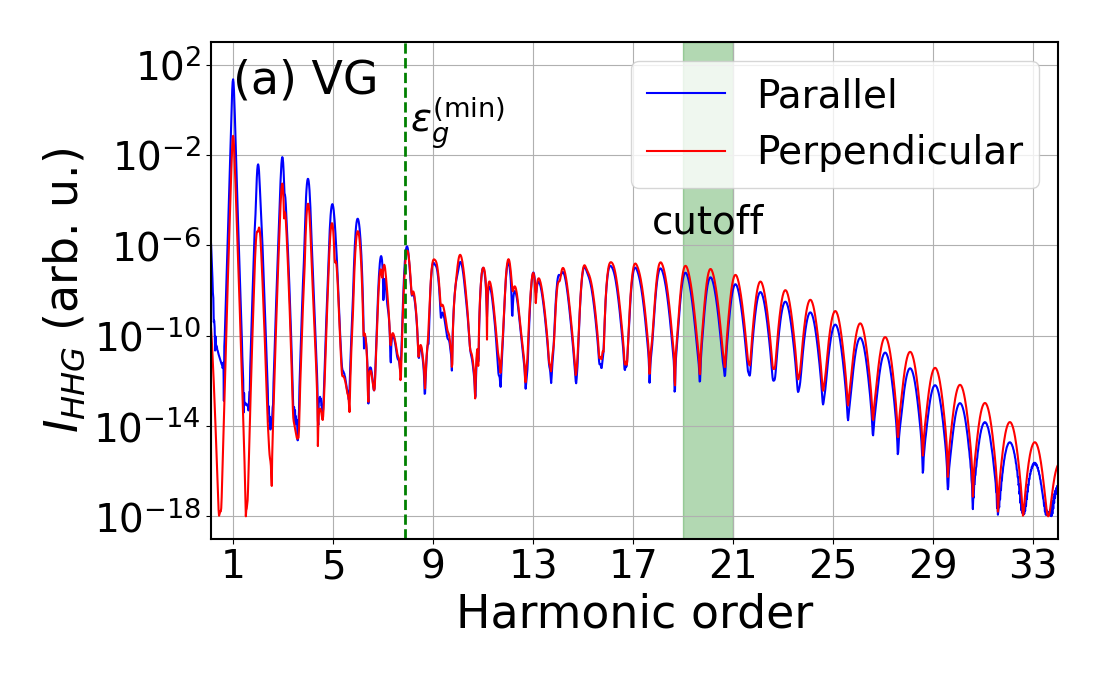} &  \includegraphics[width=0.4\textwidth]{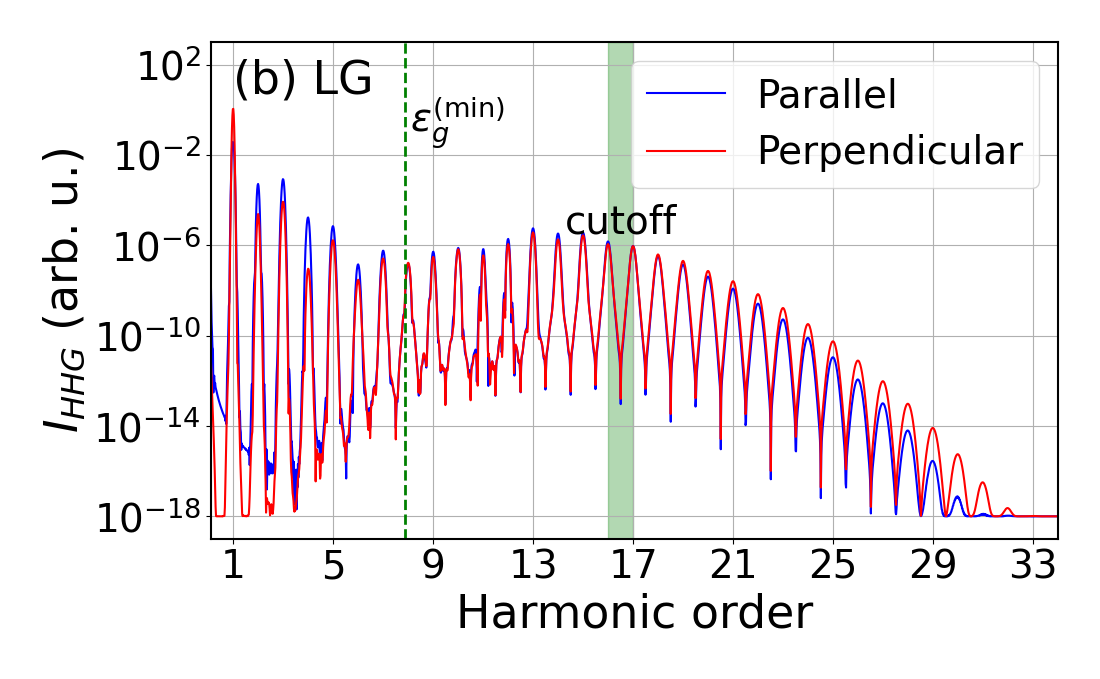} \\
        \includegraphics[width=0.4\textwidth]{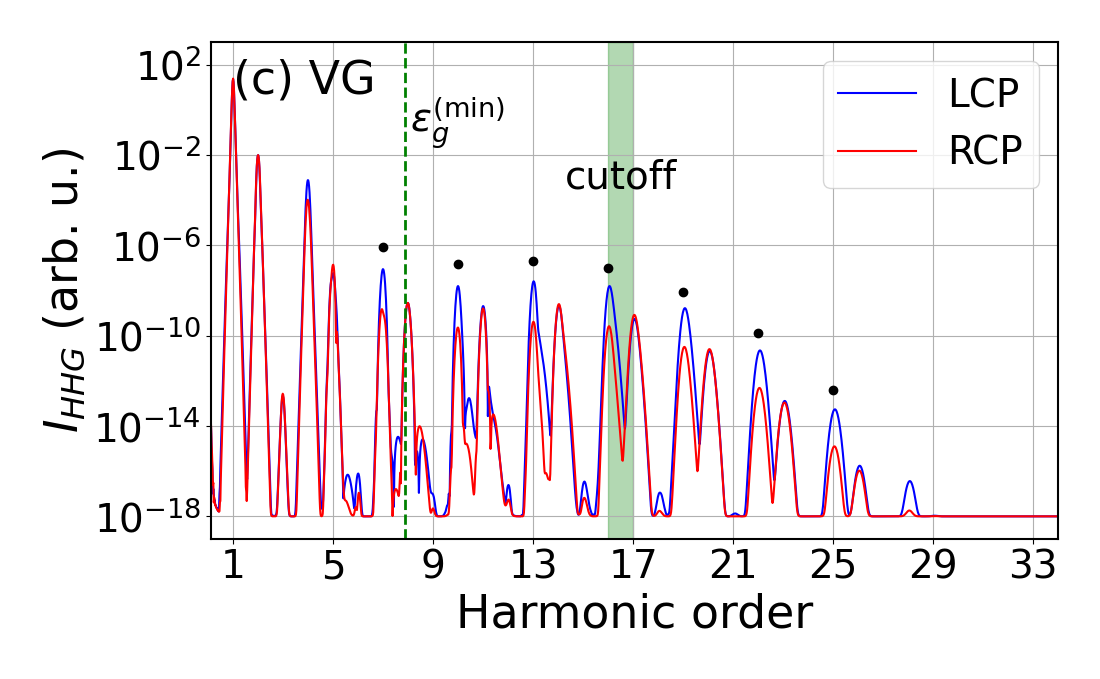} &  \includegraphics[width=0.4\textwidth]{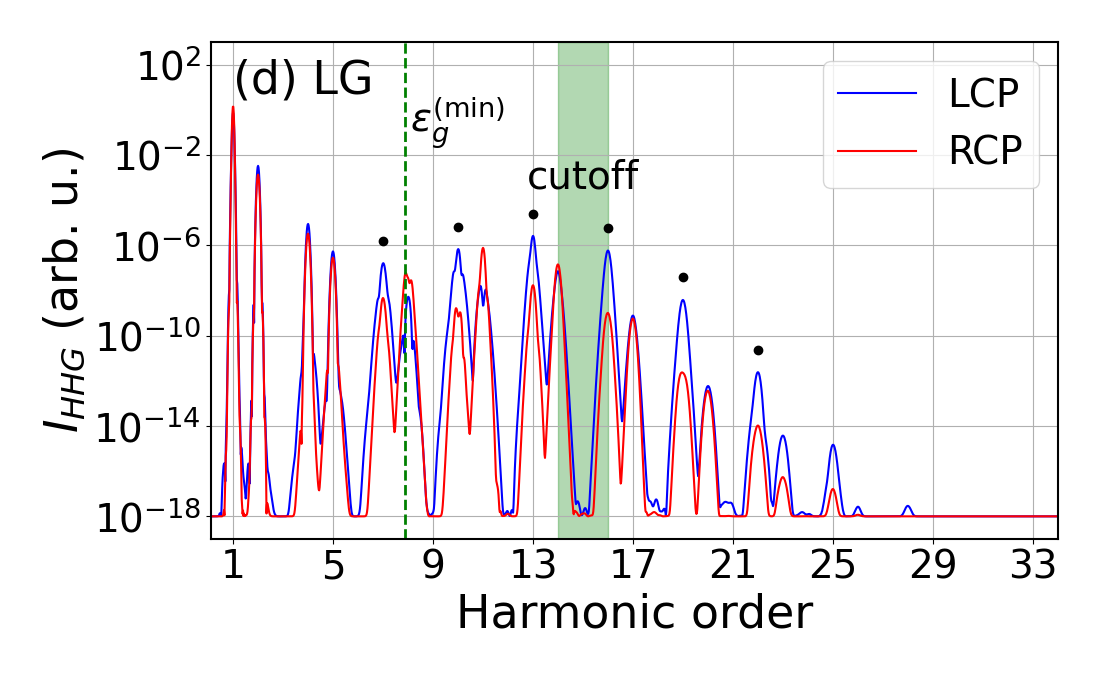}
    \end{tabular}  
    \caption{Selection rules of the HHG spectra for the VG and LG in Chern insulators. (a-b) Shows calculated HHG spectrum using linearly-polarized laser along $\Gamma$-K direction. Harmonic spectra along parallel or perpendicular direction relative to laser polarization are illustrated. (c-d) shows total harmonic spectra, $I_{\rm HHG}(\omega) = \omega\left(|J_x(\omega)|^2+|J_y(\omega)|^2\right)$, produced by circularly-polarized lasers. Results from the right-hand/left-hand circularly-polarized laser (RCP/LCP) are shown. Laser parameters given by: central frequency $\hbar\omega_{0}$ = 0.38 eV, peak electric field $E_{0}$ = 0.0045~a.u. and FWHM duration of 14 opt.~cycles under a gaussian envelope. For dephasing time $T_{2}= 5.3$~fs is used. The cut-offs of the HHG spectra are on green shadows. The black dots indicate co-rotating harmonics for the calculation of CD. The vertical dashed lines show the bandgap of the topological material.}
    \label{fig:selection_rule}
\end{figure*}

The total HHG spectra produced by linearly-polarized MIR laser for the LG and VG are in Fig.~\ref{fig:TopoGK}.~We can qualitatively find good agreement between the HHG spectrum produced by the VG and LG. Surprisingly, even in this two-band toy model,  our VG approach can reproduce the key features of HHG in TMs. 
%In the HM, the time-reversal symmetry and the inversion symmetry are broken. Thus, both even and odd harmonic orders (HOs) will emerge in the high harmonic spectrum~\cite{DenitsaPRA2021}. 
In particular, the selection rules produced by the VG for the HOs in (i) the perturbative region (low HOs of the HHG-spectra), (ii) the plateau (middle part of the HHG spectra), and (iii) the cut-off (HO to which the subsequent photon-energies drastically decrease its intensity yield) show good agreement with results from the LG (see Fig.~\ref{fig:selection_rule}(a) and (b)).

In Fig.~\ref{fig:selection_rule}, a detailed comparison between the HHG spectra in the VG and the LG is performed. Since time-reversal symmetry and inversion symmetry are broken in the HM, even harmonics and odd harmonics can be seen in the HHG spectrum~\cite{DenitsaPRA2021} along directions both perpendicular and parallel to laser polarization.  This result is gauge-symmetric, appearing both in the LG and the VG, as can be seen in Figs.~\ref{fig:selection_rule}(a) and \ref{fig:selection_rule}(b). 
%These selection rules appear in our VG approach. For example, the HHG spectrum in the parallel direction with respect to the linearly-polarized laser direction shows the even HO and odd HO structures  -- the same tendency, for the HHG spectrum along the perpendicular direction.~This well-known conclusion is shown in Figs.~\ref{fig:selection_rule}(a) and \ref{fig:selection_rule}(b) for VG and LG, respectively.~The tendency of the even and odd harmonic symmetry of the LG is ``perfectly" reproduced by our VG. This qualitative result preserves the gauge-symmetry. 
%These numerical tests show that the VG approach can successfully {\it integrate the numerical singularity in the Berry connection of topological materials} and reproduce key features of the high harmonic spectrum.

\noindent Another interesting test for the VG approach is the calculation of the Circular Dichroism (CD) produced from the HHG signal of the CIs (for details, see Ref.~\cite{Chacon2020}).~The HHG spectra are produced by left-hand and right-hand circularly polarized lasers.~We define the Circular Dichroism (CD) as the normalized difference between HOs from the left circularly-polarized laser (LCP) and right circularly-polarized laser (RCP),
\begin{eqnarray}
    {\rm CD}_k = \frac{I^{k}_{\mathrm{RCP}} - I^{k}_{\mathrm{LCP}}}{I^{k}_{\mathrm{RCP}}+ I^{k}_{\mathrm{LCP}}}.
\end{eqnarray}
Note that for materials that preserves the time-reversal symmetry, such as MoS$_2$, ${\rm CD}_{k}$ is zero.

Figures~\ref{fig:selection_rule}(c)~and~\ref{fig:selection_rule}(d) show HHG spectra produced by the VG and LG formalisms. The co-rotating HOs, $k=3n+1$, produced by LCP are much larger than the co-rotating HOs produced by the RCP driver.~We observe that for all co-rotating HOs, the ${\rm CD}=-1$, for HM parameters with $\nu=-1$. 
This is observed for both the VG and LG pictures, and is consistent with the previously reported physics of TMs Ref.~\cite{Chacon2020}.

%~This shows that VG calculation captures the central physical tendency of symmetries and physics of TMs Ref.~\cite{Chacon2020} and is capable of integrating the special topological wavefunction singularity.

%
\begin{figure}[htbp]
    %\centering
\hspace{-0.3cm}\includegraphics[width=0.45\textwidth]{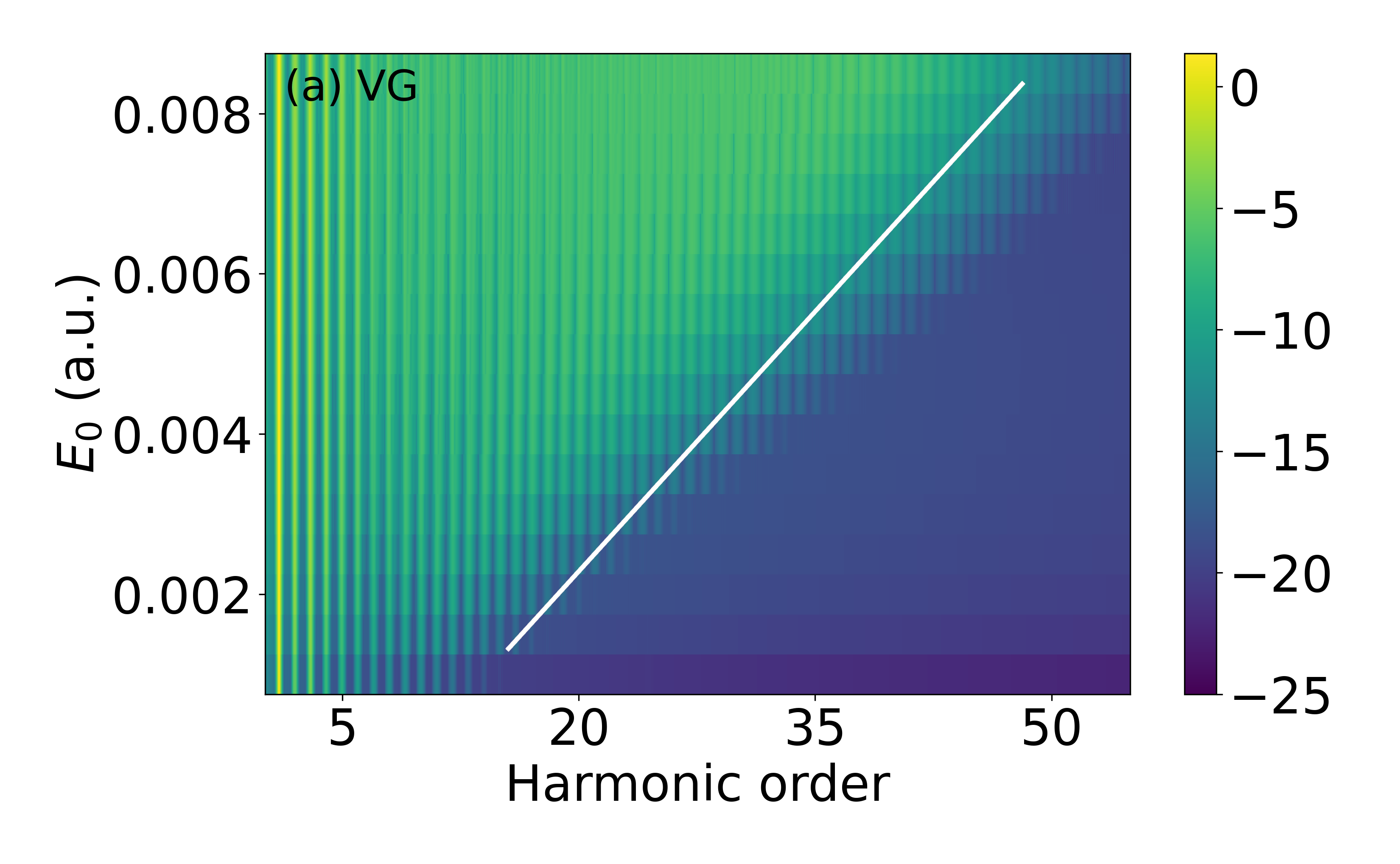}  \includegraphics[width=0.45\textwidth]{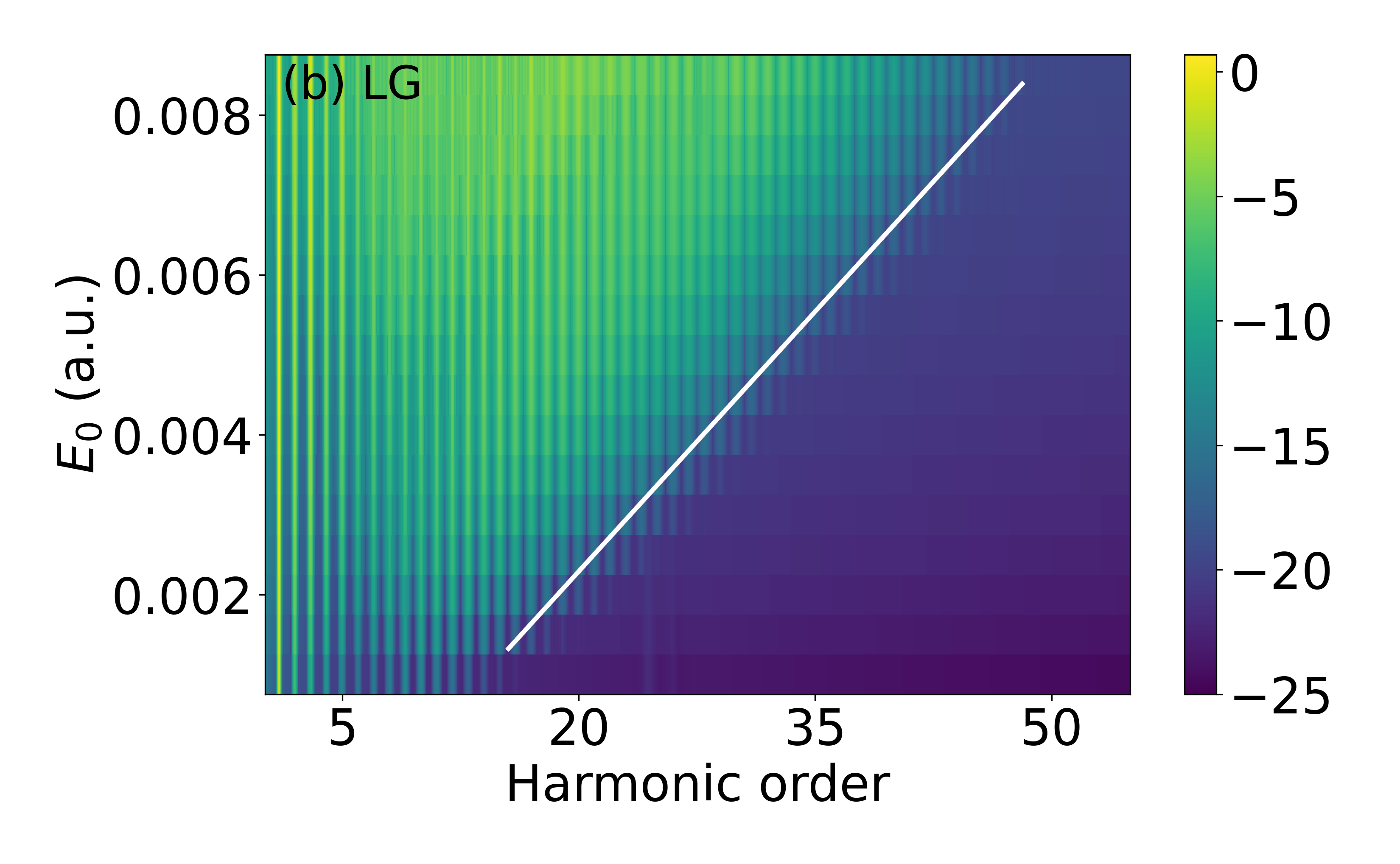}    
    \caption{Cutoff law for Chern insulators in both velocity and length gauges. Total HHG emissions, $I_{\rm HHG}(\omega)=\omega\left( |J_x(\omega)|^2 + |J_y(\omega)|^2\right)$ as a function of electric field peak strength $E_{0}$ for (a) VG and (b) LG for linearly polarized light. Other laser parameters are the same as in Fig.~\ref{fig:selection_rule}. Both velocity and length gauges show linear scaling of the high harmonic cutoff with peak electric field.}
    \label{fig:linear_scale}
\end{figure}

We now check whether the cut-off linear scaling law of the HHG spectrum can be verified within the VG approach ~\cite{GhimireNatPhy2011}. 
%Indeed, we validate that the linear cut-off for TMs using our VG approximation~\cite{GhimireNatPhy2011} (see Fig.~\ref{fig:linear_scale}). %We can then ensure that the VG and LG follow the same tendency in a broad region of the HHG spectra. 
The Harmonic spectra as a function of electric field peak strength are shown in Fig.~\ref{fig:linear_scale}. Both the VG and LG show a similar linear cut-off law: the cut-off of HHG spectra as a function of the electric field strength $E_0$ is a straight line.
\begin{figure}[htbp]
    \centering
    \begin{tabular}{c}
        \includegraphics[width=0.45\textwidth]{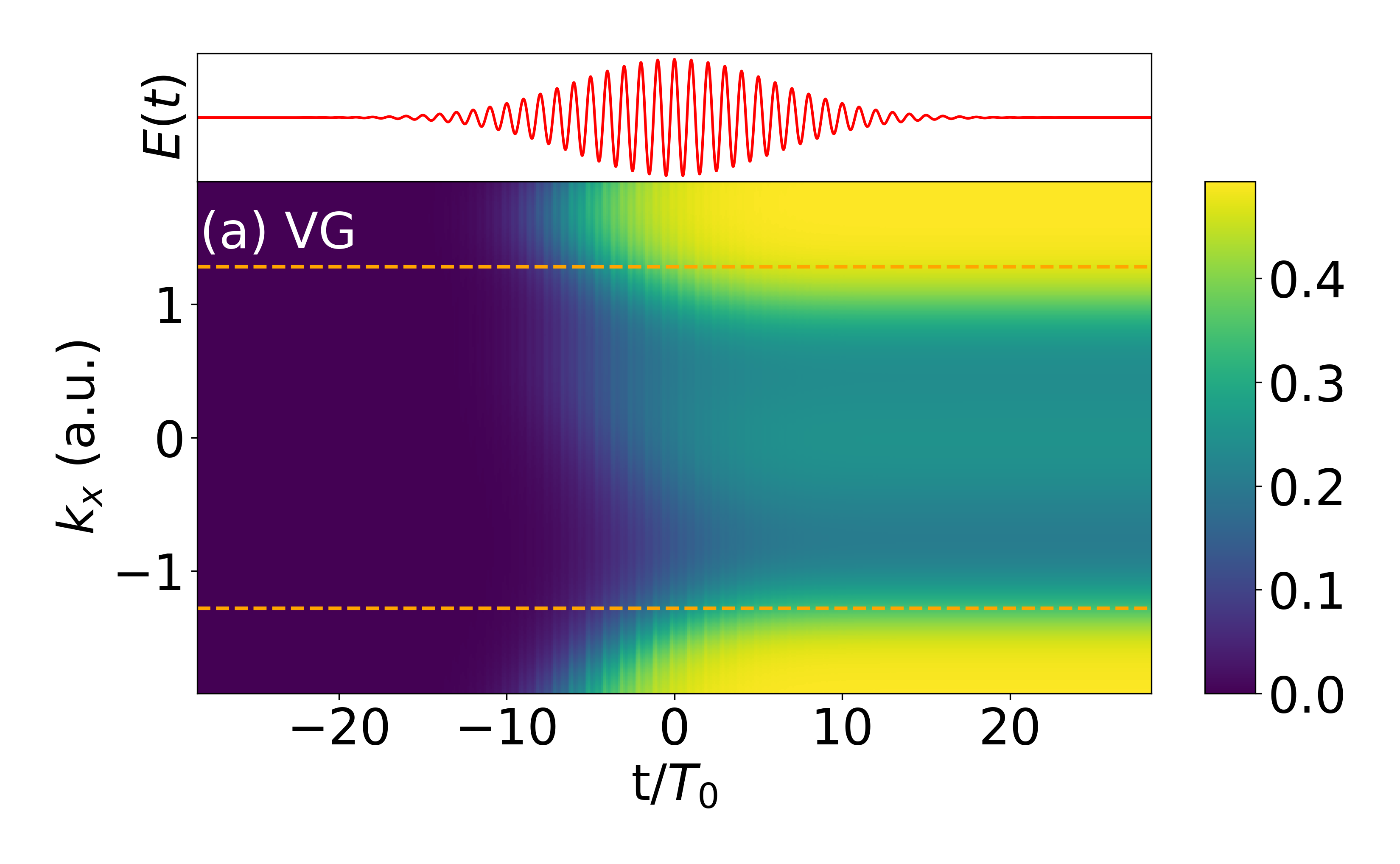} \\  \includegraphics[width=0.45\textwidth]{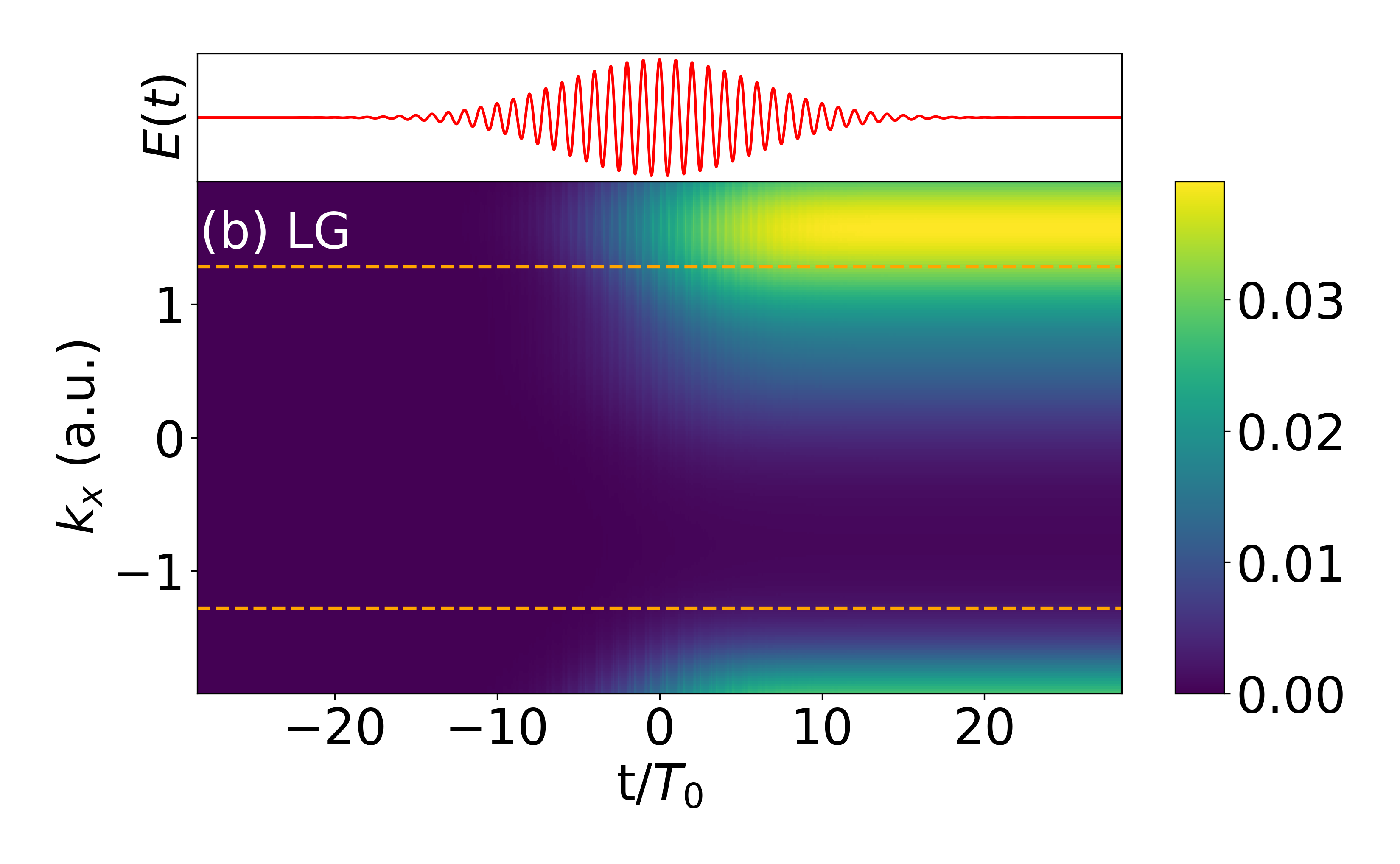}
    \end{tabular}
    \caption{Time-evolution of the ``cross-section" in the conduction band occupation. Occupation integrated along $k_{y}$ is shown as a function of time. $k_{x}$ axis is along $K'-\Gamma-K$ direction~(see Fig.~\ref{fig:bandgap}), and orange dotted line indicates the $K$ and $K'$ point in Brillouin zone. }
    \label{fig:occupa}
\end{figure}
Finally, we show the occupation of the conduction band as a function of time in Fig.~\ref{fig:occupa} around $k_y=0$ (see Fig.~\ref{fig:bandgap} for a plot in $k$-space). 

%These calculations show surprising results. One expects that the HHG spectra should be proportional to the occupation of the conduction band~\cite{VampaJPB2017,Chacon2020}, but this is not what our calculations of VG vs LG shows. LG.~Figures~\ref{fig:occupa} manifest the largest occupation propagates around K/K' points of the BZ for VG compared to LG. We hypothesize that the HHG spectra need to consider the phase provided by the coherence or off-diagonal of $\hat{\rho}({\bm k,t})$. Additionaly, there is no reason to consider that this simple two-band model will follow the VG and LG gauge-symmetry for the occupations, as it is in the case of gases~\cite{Granados2012,JAHAndPlaja2009}.
% SASHA:  I think, if you want to include the above paragraph, further analysis is necessary, since it raises more questions than it answers.

%We partially conclude that in the few band approach, that the VG can integrate the topological wavefunction and the singularity in the Berry connection successfully. However, the analysis must be carefully considered in terms of the numbers of bands of the physical material.

%We conclude that the VG approach can successfully {\it integrate the numerical singularity in the Berry connection of topological materials} and reproduce key features of the high harmonic spectrum.  
%However, the analysis must be carefully considered in terms of the numbers of bands of the physical material.

\section{Conclusions}\label{sec:D00}
We conclude that the laser-electromagnetic velocity gauge approach can successfully {\it integrate the numerical singularity in the Berry connection of topological materials} and reproduce key features of the high harmonic spectrum. 
%In this paper, we addressed whether or not the laser-electromagnetic velocity gauge can simulate the High Harmonic Generation (HHG) spectra from topological materials and ``integrate" the singularity of the Berry connection or the phase of the topological wavefunction.
Using a toy two-band model, we show that the velocity gauge can qualitatively capture the charge current and the HHG spectra without any artificial noise introduced by the singularity of the transition matrix elements, either dipole, Berry connection, or momentum. 

Additionally, we compare our results: 
(1) HHG spectra produced from linearly and circularly polarized lasers,
(2) the Circular dichroism,
(3) the linear cut-off law; to those produced by the length gauge (LG) in the maxima localized Wannier basis. 
We find good qualitative agreement in the high harmonics spectrum between the VG and the LG, both in trivial and topological materials.
%HHG spectra produced by both the VG and LG in topological materials.
The lack of quantitative agreement between the two approaches is partly due to the limited number of bands and the tight-binding approach, which we used as a proof of concept. 

We expect the VG approach to be more rigorous for TMs, since it treats the numerical singularity present within the LG approach.  Hence the velocity gauge approach presented here introduces new theoretical tools in investigating the highly nonlinear optical emission from topological materials.  

%Our velocity gauge approach opens a new avenue in understanding of the high non-linear optical emission from topological materials, and its ultrafast dynamics is an open question yet to be investigated in our future works.

\begin{acknowledgments}
	We thank professor Angel Rubio's group for calculating the TDDFT result for $\mathrm{MoS}_{2}$. D.K., D.E.K and A.C. acknowledge support by Max Planck POSTECH/KOREA Research Initiative Program [Grant No 2016K1A4A4A01922028] through the National Research Foundation of Korea (NRF) funded by the Ministry of Science, ICT \& Future Planning, partly by the Korea Institute for Advancement of Technology (KIAT) grant funded by the Korea Government (MOTIE) (P0008763, The Competency Development Program for Industry Specialist). D.S. is supported by Alexander von Humboldt Foundation. A.S.L acknowledges support by the Institute for Materials Research (IMR) at Ohio State University and Center of Emergent Materials (CEM) supported by NSF, grant No. DMR-2011876.	
\end{acknowledgments}

%\newpage

\section*{Appendix}
\appendix

\section{The Haldane model} \label{sec:Haldane}
%{\color{red} Dasol, Haldane model should be Appendix A and NO B.}

The Haldane model (HM) \cite{Haldane1988} is the first model representing the quantum anomalous Hall effect (QAHE) introducing local magnetic flux. This model is a minimum of a two-band toy model but captures the most relevant physics of the Chern insulator.~The HM considers a TBA Hamiltonian in a hexagonal lattice and hopping parameters up to the next-nearest neighborhood (NNN). 

This model can be a Chern insulator or a trivial insulator, depending on its parameter.

\subsection{Haldane's Hamiltonian}
The Haldane model is a two-band approximation obtained from a hexagonal lattice of two sub-lattices with atoms A and B. Thus, after applying the TBA for on-site potentials, the nearest-neighbor~(NN) and  the next-to-nearest-neighbor~(NNN), and changing the Hamiltonian elements from Wannier function to Bloch basis, we find,
\begin{align}
	H_{0}(\bm{k}) = B_{0}(\bm{k}) I + \bm{B}(\bm{k}) \cdot \bm{\sigma},
	\label{eq:HaldaneHamiltonian}
\end{align}
here, $I$ is the identity matrix and $\bm{\sigma}=\{\sigma_x,\sigma_y,\sigma_z\}$~Pauli's matrices.~Additionally, the $\bm{B}(\bm{k}) = \{B_{1}(\bm{k}), B_{2}(\bm{k}), B_{3}(\bm{k})\}$ is known as pseudomagnetic field. Each vector is
\begin{align}
	B_{0}(\bm{k}) &= 2t_{2}\cos\phi_{0}\sum_{i=1}^{3}\cos(\bm{k}\cdot\bm{b}_{i}), \\
	B_{1}(\bm{k}) &= t_{1}\sum_{i=1}^{3}\cos(\bm{k}\cdot\bm{a}_{i}), \\
	B_{2}(\bm{k}) &= t_{1}\sum_{i=1}^{3}\sin(\bm{k}\cdot\bm{a}_{i}), \\
	B_{3}(\bm{k}) &= M_{0} - 2t_{2}\sin\phi_{0}\sum_{i=1}^{3}\sin(\bm{k}\cdot\bm{b}_{i}), 
\end{align}
where $\bm{a}_{i}$ are the~NN vectors, and $\bm{b}_{i}$ the~NNN vectors. 

The displament vectors are given by
${\bf a}_1=\left(0,a_0\right)$, 
${\bf a}_2=\tfrac12 \left(-{\sqrt{3}},-1\right)a_0$,
${\bf a}_3=\tfrac12\left(\sqrt{3},-1\right)a_0$, 
${\bf b}_1=\left(\sqrt{3},0\right)a_0$, 
${\bf b}_2=\tfrac12\left(-\sqrt{3},+3\right)a_0$ and 
${\bf b}_3=\tfrac12 \left(-\sqrt{3},-3 \right)a_0$.

where $t_{1}$ is~the NN hopping parameter and $t_{2}$, the NNN hopping parameter. $M_{0}$ is on-site potential that breaks the inversion symmetry, and $\phi_{0}$ the local magnetic flux, which breaks the time-reversal symmetry. 

\begin{figure}
	\centering
	\includegraphics[width=0.48\textwidth]{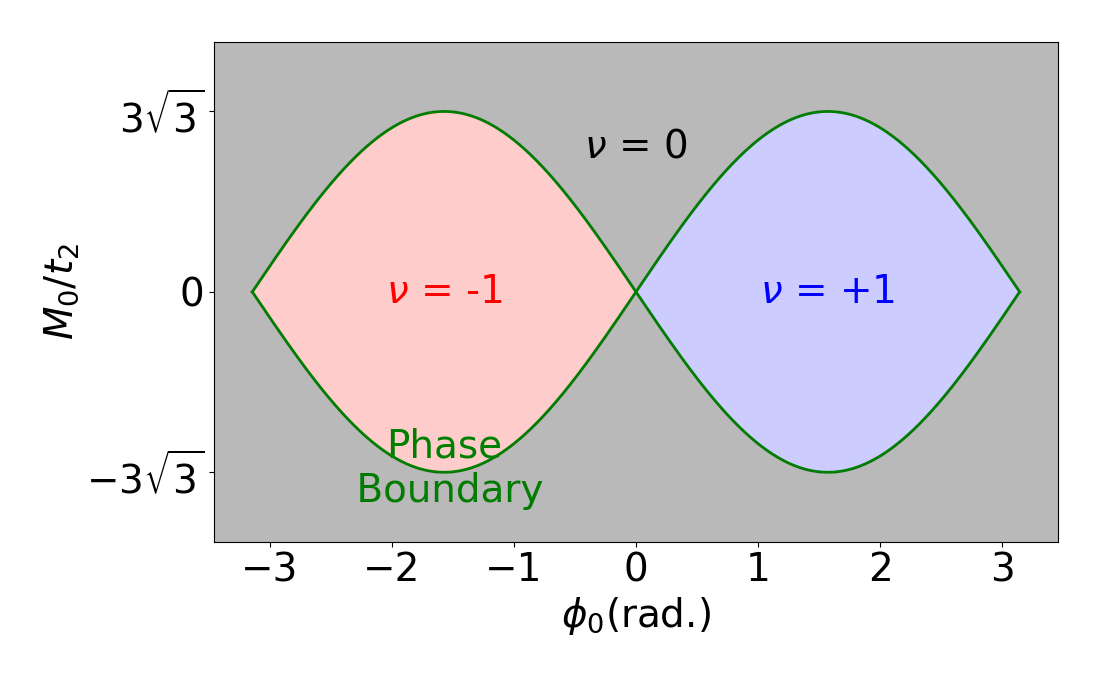}
	\caption{Topological phase diagram for the Haldane model. The diagram shows three different topological phases $\nu=\{0, \pm1\}$, where $\nu=0$ indicates topologically trivial state and $\nu=\pm1$ represents Chern insulator.}
	\label{fig:HaldanePhase}
\end{figure}

Figure~\ref{fig:HaldanePhase} shows the topological phase diagram of HM. The Haldane model yields a gapless band structure, where the topological phase transition occurs, with the condition $M_{0}/t_{2} = \pm 3\sqrt{3}\sin \phi_{0}$.

HM can have three topological invariants or Chern numbers  or topological phases $\nu=\{-1,\,0,\, +1\}$, where $\nu=0$ is trivial insulator (or ``Dirac Semimetal") and $\nu=\pm1$ is topological non-trivial phase. 

As shown in Fig.~\ref{fig:HaldanePhase}, the topological phase is determined by $\phi_{0}$ and $M_{0}/t_{2}$. $t_{1}$. These parameters affect band structure but do not affect the topological phase. By controlling those parameters, we can adjust the bandgap and topological phase to mimic a topological CI.

\subsection{Dipoles, Berry connection, Berry Curvature and Chern number}
Fortunately, we can solve 2x2 Hamiltonian analytically.~The energy dispersion of the Haldane model reads,

\begin{align}
	\varepsilon_{c/v}(\bm{k}) = B_{0}(\bm{k}) \pm |\bm{B}(\bm{k})|.
\end{align}

\begin{figure}
	\centering
	%\begin{tabular}{c c}
		\includegraphics[width=0.25\textwidth]{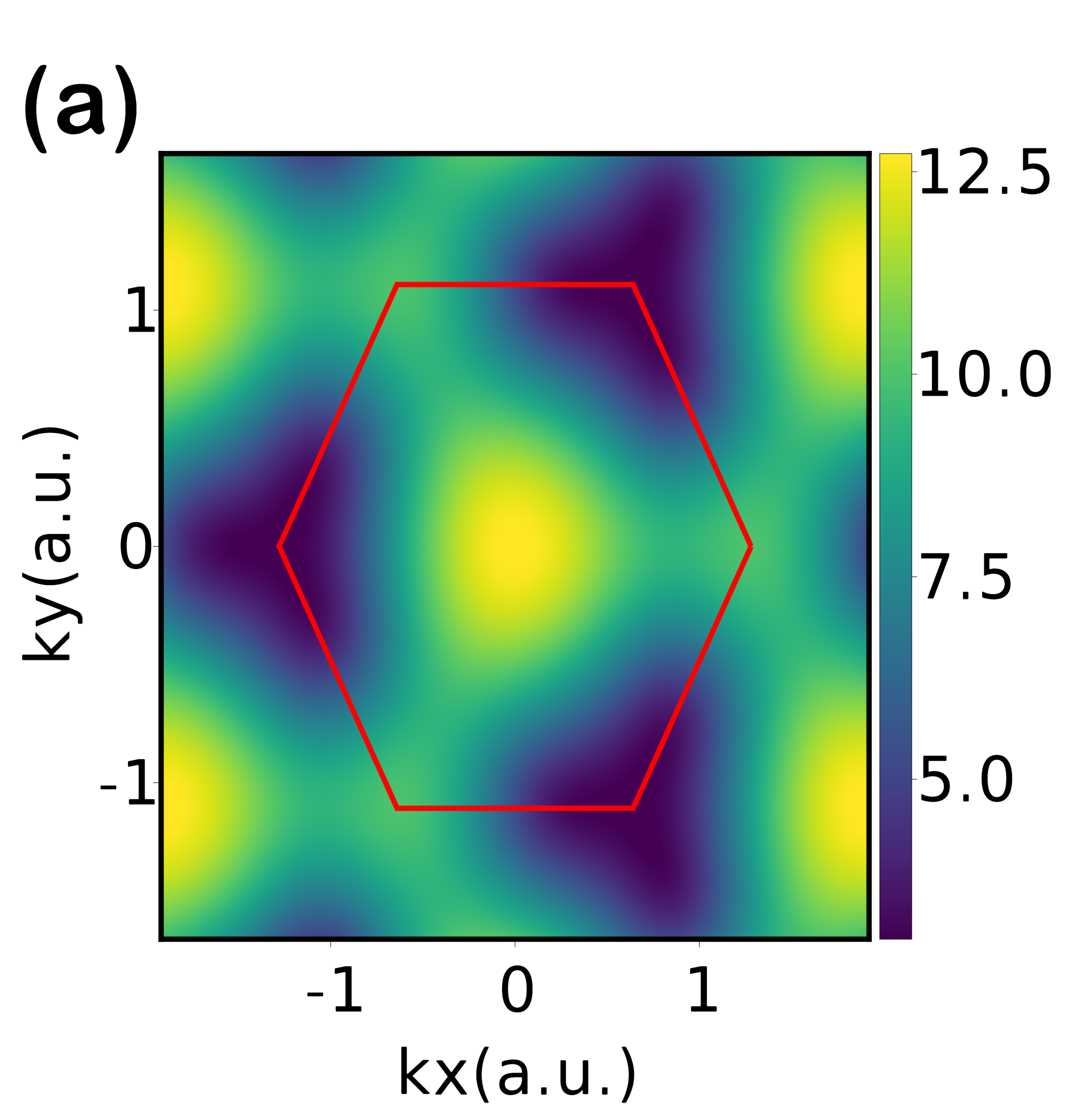} \\
		\includegraphics[width=0.25\textwidth]{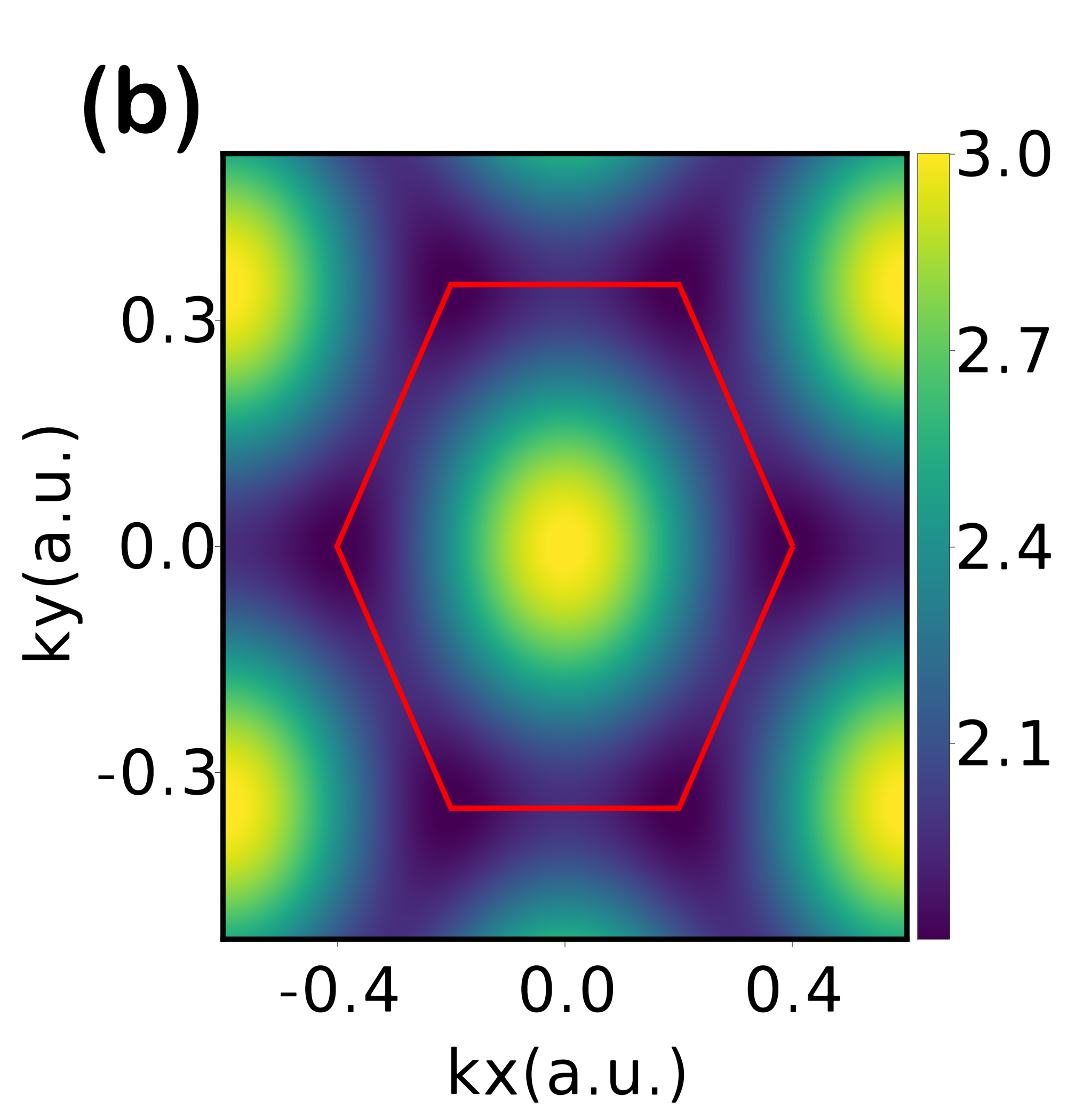}
	%\end{tabular}
	\caption{Energy dispersion for the Haldane model. Band gap (eV) for (a) Chern insulator used in the calculation and (b) trivial material, $\mathrm{MoS}_{2}$. As time-reversal symmetry is broken, (a) shows different bandgap in $K'$ and $K$ point while (b) has the same bandgap.}
	\label{fig:bandgap}
\end{figure}

The band-gap for HM is shown in Fig.~\ref{fig:bandgap}. Here we use parameters $M_{0}$ = 0.0635 a.u., $t_{1}$ = 0.075 a.u., $t_{2}$ = 0.025 a.u., and $\phi_{0}$ = 1.16 rad  for topological material. For trivial material, $\mathrm{MoS}_{2}$,  $M_{0}$ = 0.9 eV = 0.0331 a.u., $t_{1}$ = 0.4 eV = 0.0147 a.u., $t_{2}$ = 0.667 eV = 0.0245 a.u., and $\phi_{0}$ = 0 are used.

To investigate topological aspects of materials, it is required to calculate the dipole matrix elements,

\begin{align}
	{\bf d}_{m'm}({\bm k}) &= \mathrm{i}\langle u_{m',{\bm k}}| \nabla_{\bm k}|u_{m,{\bm k}}\rangle,
	\label{eq:Haldane_dipole}
\end{align}

where $\ket{u_{m}\bm{k}}$ is the periodic part of Bloch state. Usually, people distinguish diagonal and off-diagonal components in Eq.~(\ref{eq:Haldane_dipole}) and call diagonal components as Berry connection,

\begin{align}
	{\bm\xi}_m(\mathbf {k})& = {\bf d}_{mm}({\bm k})
	\nonumber \\
	& =
	\mathrm{i}\langle u_{m,{\bm k}}| \nabla_{\bm k}|u_{m,{\bm k}}\rangle.
	\label{eq:Haldane_BConnection}
\end{align}

\begin{figure}
	\centering
%	\begin{tabular}{c c}
		\includegraphics[width=0.32\textwidth]{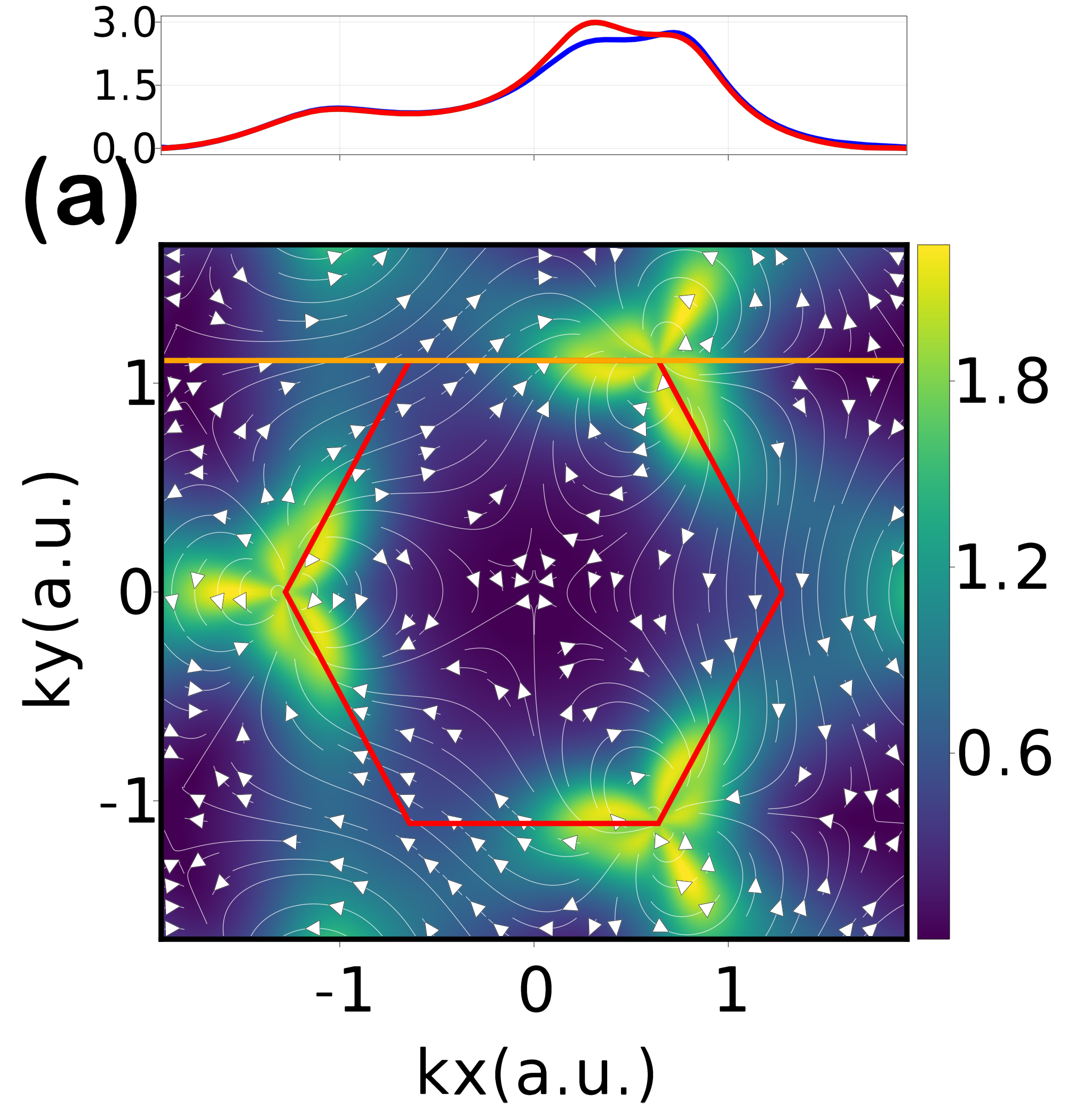} \\
		\includegraphics[width=0.32\textwidth]{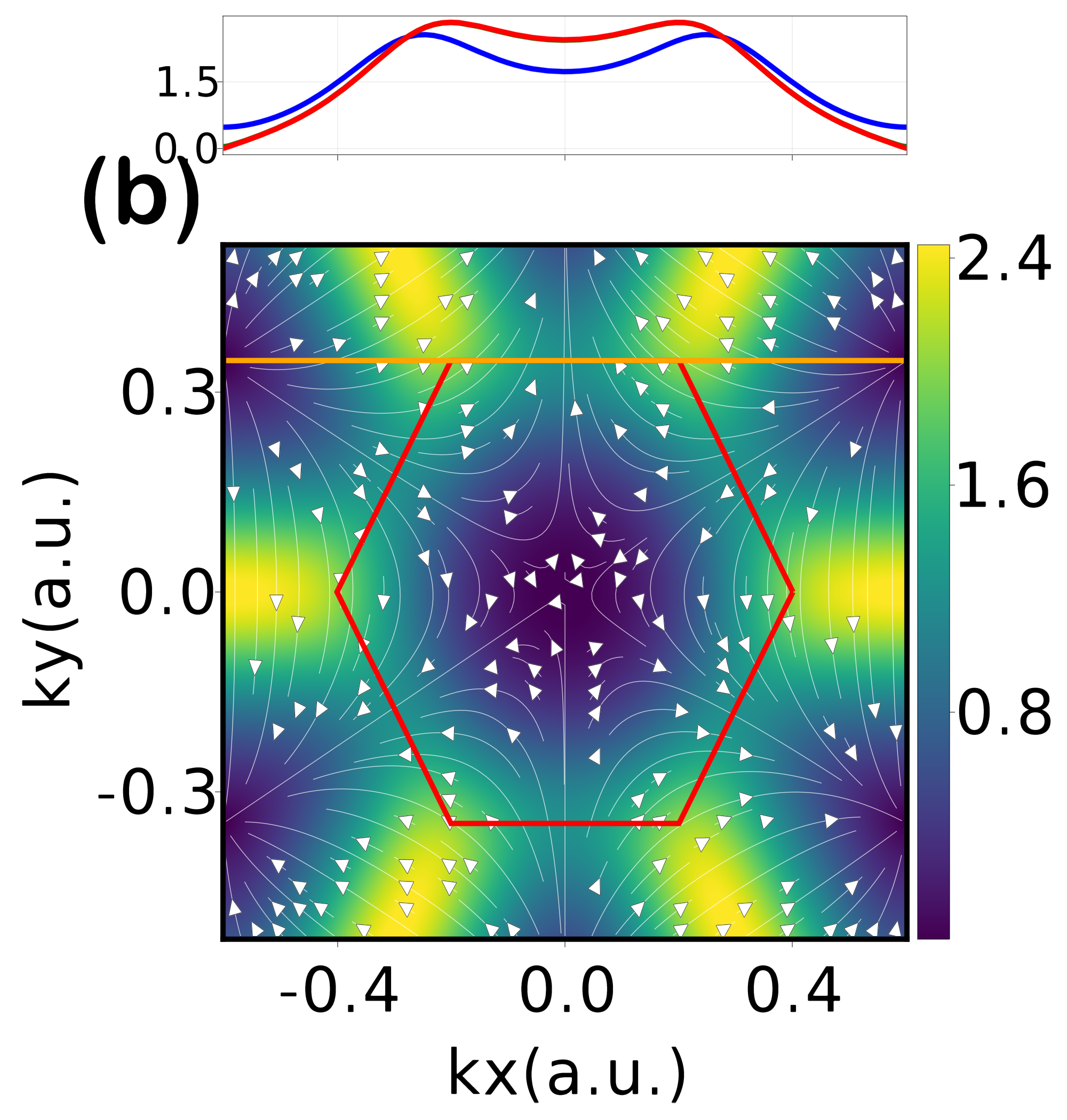}
%	\end{tabular}
	\caption{Dipole matrix element for topological materials. (a)~Real value of the dipole matrix element for Chern Insulators defined in the topological Haldane model and (b) same as in (a) but for a topologically trivial material, i.e., $\mathrm{MoS}_{2}$. The vectorial field indicates real components of the dipole matrix element. Upper panels show it cut along the orange line, i.e., the real value of dipole is plotted along the orange line with small $k_{y}$ offsets. Euclidean norm of dipole's real value is smooth for both cases, but the non-trivial topological case (a) has vortex while (b) has no discontinuity.}
	\label{fig:dipole}
\end{figure}

Berry connection and off-diagonal dipole are plotted in Figs.~\ref{fig:Figure1}~and~\ref{fig:dipole}.
The dipole matrix element shows an interesting vortex structure in topological and trivial phases, which might lead to totally different coupling with the linearly and elliptically polarized lasers (see~$K'$ points in Fig.~\ref{fig:dipole}(a) vs. Fig.~\ref{fig:dipole}(b)). 

The Berry connection has singularity while the off-diagonal dipole ``only has discontinuity" (see~Fig.~\ref{fig:Figure1}(a)). Moreover, the dipole absolute value is gauge-invariant and has no discontinuity (see Figs.~\ref{fig:dipole}(a)). Although, the Berry connection is gauge dependent the curl of the Berry connection (called as the Berry curvature):

\begin{equation}
	{\bm \Omega}_{m}({\bm k})
	= \grad_{\bm k} \times {\bm \xi}_{m}({\bm k})
\end{equation}
is gauge invariant. The integration of the Berry curvature over Brillouin zone,
%\vspace{-1cm}
\begin{equation}
	{\nu}_m
	\coloneqq
	\tfrac{1}{2\pi} \int_\mathrm{BZ} {\bm \Omega}_m(\mathbf k) \mathrm\cdot d^2\mathbf k
	,
	\label{eq:Haldane_ChernN}
\end{equation}
is a {\it topological invariant} of the system, called Chern number, which is shown in Fig.~\ref{fig:HaldanePhase}. 

\section{Gauge convariance} \label{sec:gauge}
\subsection{Sum rule for gauge covariance}

It is well-known that VG needs more bands to get convergence with LG. The main problem is that canonical commutation relation
\begin{eqnarray}
	[\hat{x}^{\alpha}, \hat{p}^{\beta}] = i\delta_{\alpha\beta},
	\label{eq:commutator_rp}
\end{eqnarray}
where $\{\alpha, \beta\} = \{x, y, z\}$, is generally not valid when we have finite bands~\cite{ventura2017,virk2007,kruchinin2013,yakovlev2017}. 

In the plane wave basis, we can numerically satisfy this relation by increasing the number of bands up to the convergence. In TBA, however, we can not state that more bands always give better results given the constrain of Eq.~(\ref{eq:commutator_rp}).

\noindent Furthermore, from the definition of $\hat{\bm{x}}$, and the Heisenberg relationship with the evolution of given operator, the momentum matrix reads, 
\begin{eqnarray}
	\bm{\hat{P}} = \frac{\partial \hat{H}_{0}}{\partial \bm{k}} - i[\bm{\hat{D}}, \hat{H}_{0}].
	\label{eq:momentum_operator}
\end{eqnarray}

If we use the Hamiltonian gauge, $H_{0}$ becomes a diagonal matrix of energy dispersion~\cite{VanderbiltPRB1997,VanderbiltPRB2006}. Then, Eq.~(\ref{eq:momentum_operator}) becomes the well-known formula~\cite{VanderbiltPRB2006,PeresPRB2011}, 

\begin{eqnarray}
	\bm{P}_{mn} = \left\{\begin{array}{cc}
		\frac{\partial }{\partial \bm{k}}\varepsilon_{m} & \mathrm{if}\,\, m=n, \\
		i(\varepsilon_{m} - \varepsilon_{n})\bm{d}_{mn} & \mathrm{if}\,\, m\neq n
	\end{array}\right\}, \label{eq:momentum}
\end{eqnarray}

in which the first term is the intra-band component, and other terms define the inter-band currents.

In Wannier representation, $H_{0}$ is modeled by the Tight-Binding Approximation (TBA) which implies:
\begin{eqnarray}
	\bm{P} = \frac{\partial H_{0}}{\partial \bm{k}} - i[\bm{D}^{\rm (W)}({\bm k}), H_{0}({\bm k})].
	\label{eq:momentum_operator}
\end{eqnarray} 

Using Eq.~(\ref{eq:momentum_operator}), Eq.~(\ref{eq:commutator_rp}) becomes sum rule,

\begin{align}
	i\delta_{\alpha\beta}I &= i\frac{\partial^{2} \hat{H}_{0}}{\partial k_{\alpha}\partial k_{\beta}} + \frac{\partial}{\partial k_{\alpha}} \left[\hat{D}^{\beta}, \hat{H}_{0}\right] \nonumber \\ 
	&+ \left[\hat{D}^{\alpha}, \frac{\partial \hat{H}_{0}}{\partial k_{\beta}}\right] + i\left[\hat{D}^{\beta}, \hat{H}_{0}\right]\hat{D}^{\alpha}.
	\label{eq:sum_rule}
\end{align}

If we use condition $D^{\rm(W)}({\bm k})={\bm 0}$, sum rule in Wannier representation becomes

\begin{eqnarray}
	\delta_{\alpha\beta} I = \frac{\partial^{2} H_{0}}{\partial k_{\alpha}\partial k_{\beta}},
\end{eqnarray}

which is only valid when $H_{mn} \approx \frac{1}{2}k^{2}\delta_{mn}$, free-electron case. This shows that sum rule is always broken with above condition, therefore, the laser-electromagnetic gauge-symmetry too. 

\subsection{Conversion between electromagnetic gauge}
We can convert operators in both gauge as \cite{ventura2017}

\begin{eqnarray}
	\hat{O}^{\rm (VG)}(\bm{k}) = \hat{O}^{\rm (LG)}(\bm{k} + \bm{A}(t)).
\end{eqnarray}

However, the form of the operator $\hat{O}^{\rm (VG)}(\bm{k})$ is straightforward; the matrix form of relationship between the density matrices of the VG and LG is complicated~\cite{ernotte2018, yue2020}.

\begin{subequations}
	\begin{eqnarray}
		\hat{O}^{\rm (LG)}({\bm k}+{\bf A}) &= R({\bm k},{ \bm A}) \hat{O}^{\rm (VG)}(\bm{k}) R^{\dagger}({\bm k},{\bm A}) , \\
		R_{mn}(\bm{k},\bm{A}) & \equiv \braket{u_{m\bm{k}+\bm{A}}|u_{n\bm{k}}}.
	\end{eqnarray}
	\label{eq:em_gauge_conversion}
\end{subequations}

In TBA, Eq.~(\ref{eq:em_gauge_conversion}) can be calculated by

\begin{eqnarray}
	R^{\bm{k},\bm{A}} = \hat{U}^{\bm{k}+\bm{A}\dagger}\hat{U}^{\bm{k}}
\end{eqnarray}
where $\hat{U}^{\bm{k}}$ is unitary eigenvector matrix of unperturbed Hamiltonian $\hat{H}_{0}$.

\section{Effects of dephasing in length and velocity gauges}  \label{sec:dephasingSec}
This appendix numerically studies the effects of the dephasing $T_2$ on the high harmonic generation (HHG) process for our topological Chern insulator.

Since already the phenomenological dephasing time, $T_2$, can be considered an external term related to the scattering and thermal processes in a medium, $T_2$ acts differently in both gauges~\cite{ernotte2018, yue2020}. 

These are the primary sources of discrepancies in breaking the laser-electromagnetic gauge-symmetry. To avoid this $T_2$ effect, one can convert their density matrix for each time step - convert VG density matrix to LG density matrix, apply dephasing time and come back to VG density matrix~\cite{yue2020}. However, this procedure slows down the calculation speed of the HHG spectra in VG, e.g. the computational numbers of operations and the time spent on this calculation. The VG SBEs becomes even slower than length gauge SBEs. Nevertheless, when this is compared with the Wannier basis, VG still has its advantage since Wannier LG also needs transformation to apply $T_2$. 

\begin{figure}
    \centering
%    \begin{tabular}{c}
        \includegraphics[width=0.4\textwidth]{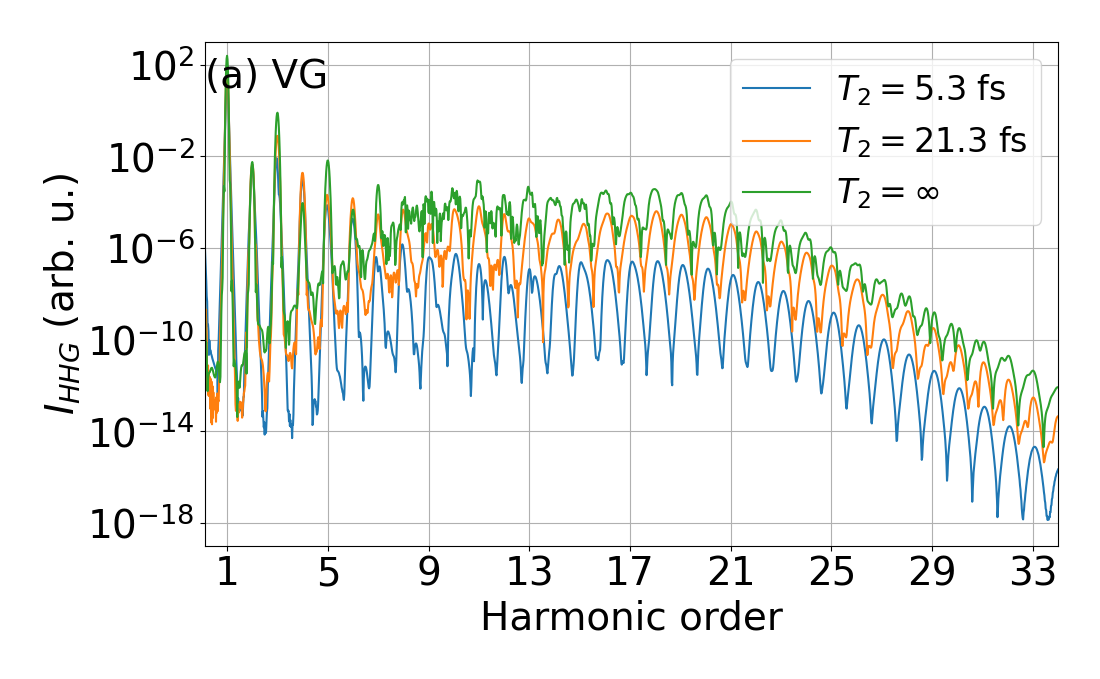} \\ \includegraphics[width=0.4\textwidth]{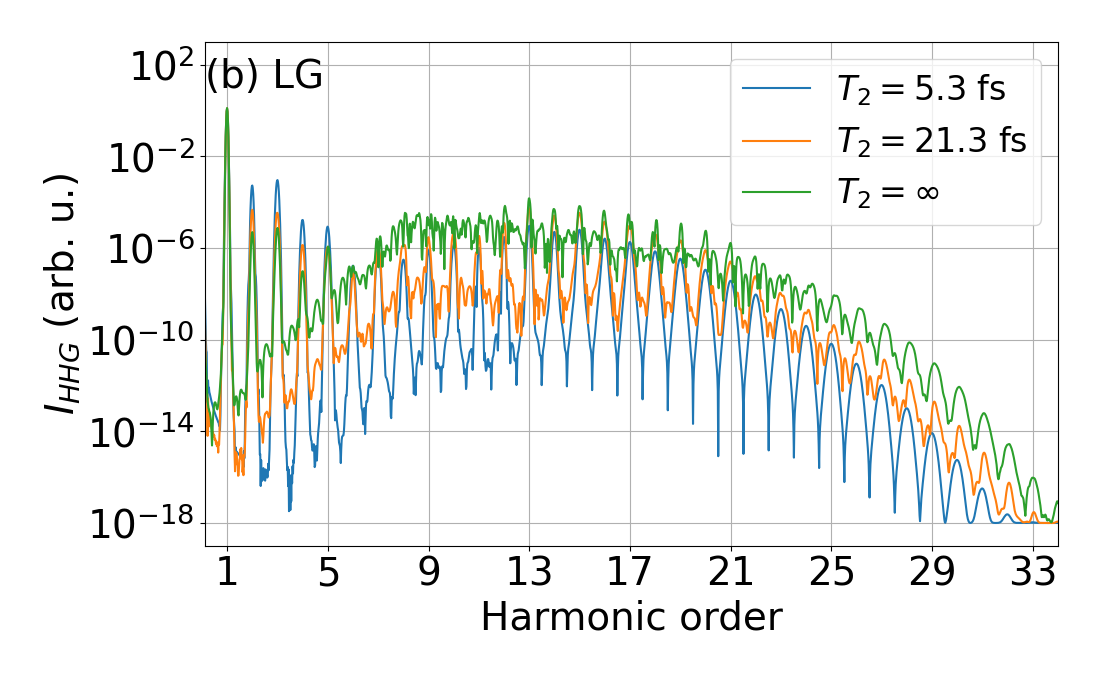}
%    \end{tabular}
    \caption{Effect of $T_2$ for Chern insulator in LG and VG. HHG calculation results for (a) VG and (b) LG is shown. Laser is linearly polarized along $\Gamma-K$ direction, and laser parameters of Fig.~\ref{fig:selection_rule} is used.} 
    \label{fig:t2scan}
\end{figure}

In Figs.~\ref{fig:t2scan}(a) and \ref{fig:t2scan}(b) we depict the HHG spectra as a function of the phenomenological dephasing time $T_2$ for a topological Chern Insulator (CI). The perturbative region of the HHG spectra $\left[1^{\rm st}, 6^{\rm th}\right]$~HOs shows different tendencies for both gauges. Significantly in the LG, the harmonic yield increases while the $T_2$ decreases for low HOs~\cite{VampaPRL2014}. In contrast, we find that the VG provides opposite behavior than LG for low orders as a function of $T_2$.

For the plateau and cut-off regions of the HHG spectra, the dephasing time $T_{2}$ induces a symmetry-gauge breaking in the spectrum of HHG too. Nevertheless, since $T_2$ in the VG and LG gauges reduce the noise in the emission signal, it is hard to quantify the difference between both gauges in Fig.~\ref{fig:t2scan}. It is more obvious to observe the differences through current. 

\begin{figure}
    \centering
    \includegraphics[width=0.4\textwidth]{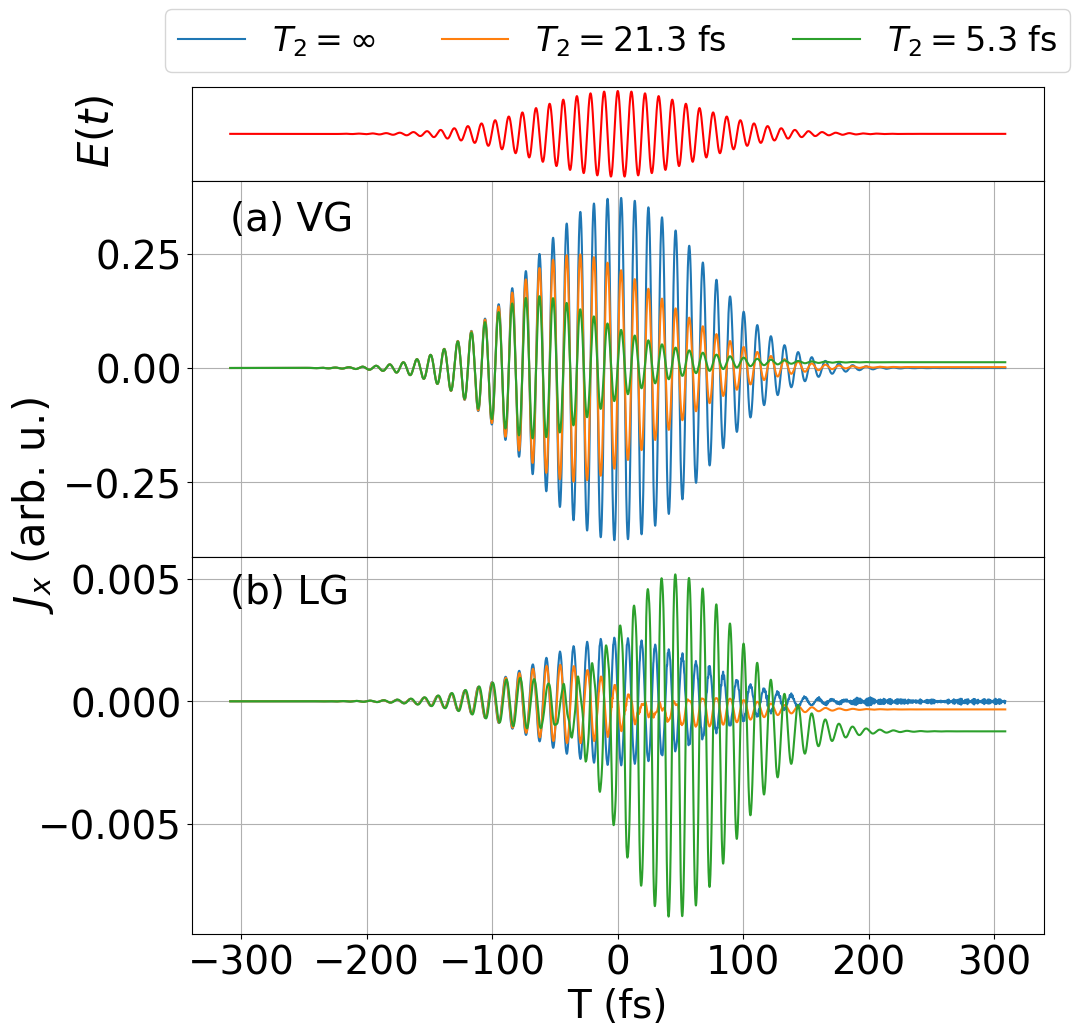}
    \caption{Effect of $T_2$ for Chern insulator in length gauge and velocity gauge in current. Total current for (a) velocity gauge and (b) length gauge are shown. All the parameters are same as Fig.~\ref{fig:t2scan}.}
    \label{fig:t2current}
\end{figure}

Figures~\ref{fig:t2current} illustrate the effect of $T_{2}$ on the currents as a function of time. For VG, they act like a window function that removes the contribution from the later time domain. For LG, the effect is more complex and global. Noise at $T_{2}=\infty$ is filtered, and envelope shape is also changed.

%\subsection{Dephasing time in the length and velocity gauges}

\subsection{Computational complexities}

We show brief illustration of computational cost for each method. In the case of LG SBEs, there are several choices like gradient or moving frame, but we will only mention the moving frame with tight binding model here. For both LG and VG SBEs, they have to calculate SBEs itself by matrix multiplication and addition and it costs about $f_{\mathrm{SBEs}}=O(N_{k}N_{b}^{3})$ by set matrix multiplication cost as $O(N_{b}^{3})$. Here, $N_{k}$ is total number of k-space grid and $N_{b}$ is number of bands. LG SBEs have to calculate matrix elements at $\bm{K}+\bm{A}(t)$ for each time step. Then it needs $f_{\mathrm{GenMatrix}}=O(N_{k}N_{b}^{2})$ order to generate appropriate matrices when assume that run time for calculate each components is $O(1)$. Then computation complexity for LG and VG SBEs without dephasing time is
\begin{eqnarray}
	f_{\LG} &= f_{\mathrm{SBEs}} + f_{\mathrm{GenMatrix}}, \\
	f_{\VG}^{\mathrm{nodeph}} &= f_{\mathrm{SBEs}}.
\end{eqnarray}
If we include dephasing time, LG SBEs have almost no additional cost , but Wannier LG SBEs and VG have conversion cost about $f_{\mathrm{dephasing}}=O(N_{k}N_{b}^{3})$ which include generating eigenvector matrix and multiply this. Real cost for dephasing time in VG and Wannier LG is little bit different, but the difference is small enough. Then computational cost for Wannier or VG is
\begin{eqnarray}
	f_{\LG}^{(W)} &= f_{\mathrm{SBEs}} + f_{\mathrm{GenMatrix}} + f_{\mathrm{dephasing}} \\
	f_{\VG} &= f_{\mathrm{SBEs}} + f_{\mathrm{dephasing}}.
\end{eqnarray}

\bibliographystyle{apsrev4-1}
\bibliography{ReferencesTo, WHComparison,references}

\end{document}